\documentclass[useAMS,usenatbib,referee]{mn2e}
\usepackage{graphicx}
\usepackage{amssymb}

\def    \SIG     {$\mathcal S$}
\def    \A       {$\mathcal A$}
\def    \HS      {$\mathcal H_S(S)$}
\def    \HA      {$\mathcal H_A(A)$}

\def    \ang     {\AA \ }

\def    \degt    {\,\, ^{\circ}}

\def \nights        {95 }

\def \imnum         {617 }

\def \directaim     {286 }
\def \directbim     {177 }
\def \directcim     {154 }

\def \objnum        {6487 } 

\def \objnumf       {6418 } 

\def \allmeas       {899,509 }

\def \allvar        {290 } 

\def \allpernum     {113 } 

\def \nonpernum     {177 } 
\def \knownnonper   {78 }
\def \nonpernew     {99 } 

\def \ebnum         {12 }
\def \knowneb       {7 }
\def \ebnew         {5 } 

\def \cephnum       {45 } 
\def \knownceph     {37 } 
\def \cephnew       {8 } 
\def \cephnewvar    {5 } 
\def \cephnewclass  {3 } 
 
\def \pernum        {56 } 
\def \knownper      {30 } 
\def \pernew        {26 } 
\def \pernewvar     {19 } 
\def \pernewclass   {7 }  

\def \combnumtotal      {59 } 
\def \combnum           {50 } 
\def \badcomb           {9 } 

\title[Wise Observatory M33 $V$-Band Variability Search]{Long-Term V-Band Monitoring of the Bright Stars of M33 at the Wise Observatory}
\author[Shporer \& Mazeh]{A. Shporer$^1$\thanks{E-mail: shporer@wise.tau.ac.il}
 and  T. Mazeh$^1$\\
  $^1$Wise Observatory, Tel Aviv University}
\date{Released 2005 Xxxxx XX}

\pagerange{\pageref{firstpage}--\pageref{lastpage}} \pubyear{2005}

\def\LaTeX{L\kern-.36em\raise.3ex\hbox{a}\kern-.15em
    T\kern-.1667em\lower.7ex\hbox{E}\kern-.125emX}

\begin{document}

\label{firstpage}

\maketitle

\begin{abstract}

We have conducted a long-term $V$-band photometric monitoring of M33 on
\nights nights during four observing seasons (2000 -- 2003). 
A total number of \objnumf lightcurves
of bright objects in the range of 14 -- 21 mag have been
obtained. All measurements are publicly available. A total of 127 new
variables were detected, of which 28 are periodic.
Ten previously known non-periodic variables were identified as 
periodic, 3 of which are Cepheids, and another previously known periodic
variable was identified as an eclipsing binary.
Our derived periods range from 2.11 to almost 300 days.
For \combnum variables we
have combined our observations with those of the DIRECT project,
obtaining lightcurves of up to 500 measurements, with a time-span of
$\sim\!$~7 years.
We have detected a few interesting variables, including a 99.3 day
periodic variable with a 0.04 mag amplitude, at the position of SNR 19. 
\end{abstract}


\section{Introduction}
\label{sec:intro}

Systematic searches for variable objects in M33 started with the seminal
work of \cite{Hubble26}, who found 42 variables in this galaxy. He
identified 35 of the variables as Cepheids, with which he established
the extragalactic nature of M33.  A few additional studies have been
conducted since then (\citealt{Hubble53}, \citealt{VanDenBergh75},
\citealt{Sandage83}, \citealt{Kinman87}) using photographic 
techniques. 

The first CCD-based variability search in M33 was conducted as part of the
DIRECT project (\citealt{Kaluzny98}, \citealt{Stanek98})
with the goals of finding Cepheids and Eclipsing Binaries (hereafter EBs) 
that will yield a
better distance estimate to our neighboring galaxies (M31 \& M33). 
The first DIRECT observational campaign of M33 (\citealt{Macri01a,Macri01b}) 
was carried out with the Whipple Observatory 1.2-m telescope
and with the Michigan-Dartmouth-MIT Observatory 1.3-m
McGraw-Hill telescope. Three 10.8' $\times$ 10.8' fields were selected
to cover the central part of the galaxy, labeled M33A, M33B and M33C,
located North, South and South-West of the center, respectively.
Observations during 42 nights, from September 1996 to October 1997, 
revealed 544 variables, including 251 Cepheids, 47 EBs
and 62 unclassified periodic variables, in the M33A and
M33B fields (\citealt{Macri01b}).

The DIRECT second observational campaign
(\citealt{Mochejska01a,Mochejska01b}) aimed to follow two detached
EBs in M33A and M33B. Observations were conducted with
the Kitt Peak National Observatory 2.1-m telescope, using 10.4'
$\times$ 10.4' FOVs centered on the central coordinates of the M33A
and M33B fields of \cite{Macri01b}. Two separate runs, each of
7 nights, were conducted in October and November 1999. 
A total of 892 new variables were detected in both
fields, increasing the number of DIRECT variables to 1436.

In this study we wished to continue the DIRECT thorough search for
variable stars in M33, with an emphasis on long time span, so we could
identify variables with longer time scales, and in particular stars
with double periodicity. We therefore performed a photometric
monitoring of the same three fields with the 1-m~telescope at the Wise
Observatory on \nights nights from September 2000 until November 2003.

Ideally, we would combine our data with the individual measurements of
the DIRECT project and search the combined data for new variables.
However, the DIRECT individual measurements were publicly available only
for the variables identified in M33A \& M33B. Therefore we could
combine the data only for the stars already identified by DIRECT as
variables. We did that in order to get a better coverage of the
periodicity and to try and look for additional modulations. For all
the other stars we performed periodicity and variability searches
independent of the DIRECT data. Although the precision of our
measurements were inferior to those of the DIRECT project, the long
time span enabled us to detect a few unknown variables, some of which
are quite interesting. Among those are an optically-periodic X-ray source
and a periodic variable at a SNR position. 

Observations and data processing are described in \S~\ref{sec:data_proc}. 
Section~\ref{sec:temp_anal} describes the 
temporal analysis techniques and results. It also describes the method of 
combining our data with the DIRECT data and compares our results with the DIRECT. 
We present a few interesting variables in \S~\ref{sec:int_var} and state concluding 
remarks in \S~\ref{sec:conclusions}.

\section{Observations and Data Processing}
\label{sec:data_proc}

\subsection{Observations}
\label{subsec:observations}

We observed M33 with the 1-m telescope at the Wise Observatory 
 from September 2000 until November
2003. Observations were carried out with a standard Cousins-Johnson
$V$ filter using a Tektronix 1024$\times$1024 back-illuminated CCD,
with a pixel scale of $0.696~\pm~0.002$~``/pixel and an
11.88'$\times$11.88' overall field of view \citep{Kaspi99}. Exposure
time was 900 seconds.  

\begin{figure*}
\includegraphics[width=13cm,height=10cm]{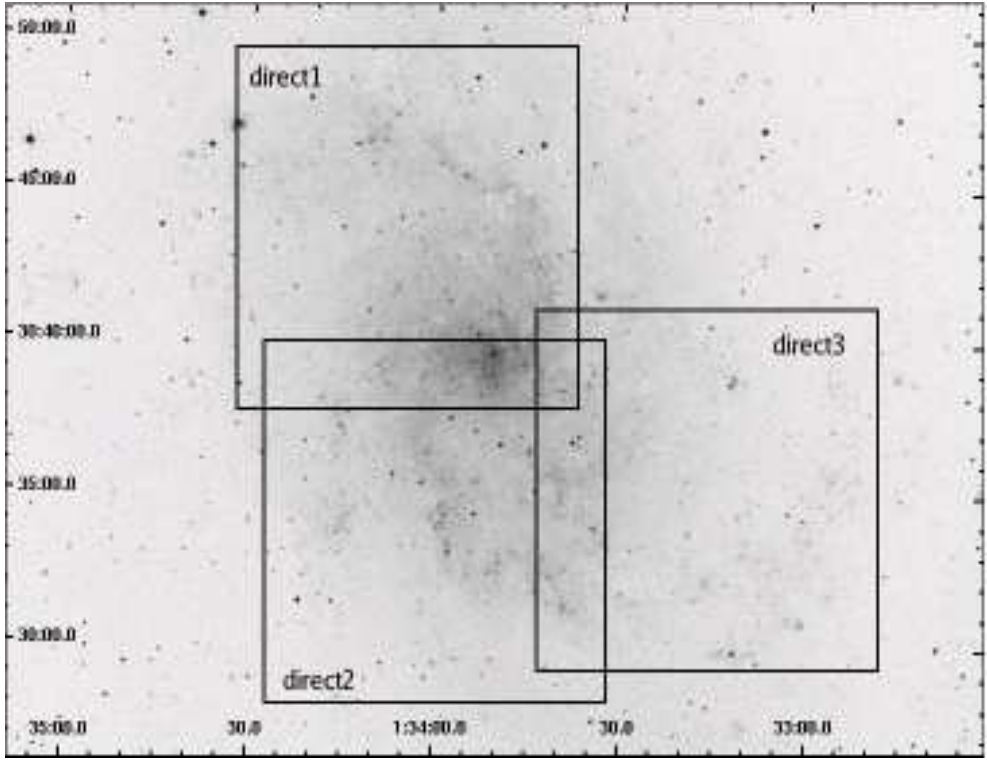}
\caption{Boundaries
of the three fields over-plotted on a Palomar Quick-$V$ survey image of M33.
North is up and East is to the left. Coordinates are in the J2000.0 system.}
\label{fig:fields_boundaries}
\end{figure*}

M33 central region was covered by three fields --- \emph{direct1}, \emph{direct2} 
and \emph{direct3}, similar to the DIRECT three fields.
Table~\ref{table:fields_coords} lists the fields central coordinates, and
Fig.~\ref{fig:fields_boundaries} shows the field boundaries on a Palomar 
Quick-$V$ survey image of M33\footnote{The compressed files of the Space Telescope Science Institute Quick-Survey of the northern sky are based on scans of plates obtained by the Palomar Observatory using the Oschin Schmidt Telescope.}.

During the three years of the project M33 was observed on \nights
nights, with about 30 nights per typical observing season (In the season of 
2002 -- 2003 M33 was observed only on 3 nights). 
We made an effort to observe each of the three fields twice per
observing night, although some technical and/or weather conditions 
did not allow all six exposures to be acquired on all nights. 
A total of \imnum
exposures were obtained and are listed on Table~\ref{table:images_jd}. They 
include \directaim$\!\!$ exposures of \emph{direct1}, \directbim of
\emph{direct2} and \directcim of \emph{direct3}.  

On the night of Aug. 31, 2003, at the last stage of our monitoring of M33, a
nova eruption was discovered in the \emph{direct1} field
\citep{IAUC8195}. To follow-up the nova evolution, we obtained 111 
exposures for the \emph{direct1} field between 
Sep. 7 and Sep. 22, 2003 \citep{IAUC8199}. Those images are included in our 
present analysis.

\begin{table*}
\begin{minipage}{60mm}
\caption{Fields central coordinates (J2000.0)}
\label{table:fields_coords}
\begin{tabular}{ccc}
\hline
Field name & RA              & Dec            \\ \hline
direct1    &   01:34:05.1    &   +30:43:43    \\
direct2    &   01:34:00.0    &   +30:34:04    \\
direct3    &   01:33:16.0    &   +30:35:15    \\ \hline
\end{tabular}
\end{minipage}
\end{table*}

\begin{table*}
\begin{minipage}{120mm}
\caption{Mid-exposure JD of all \imnum $V$ exposures obtained in the course 
of this survey.}
\label{table:images_jd}
\begin{tabular}{cccccccc}
\hline
& & & \multicolumn{2}{c}{direct1} & & & \\
\hline
  1801.290&    1896.308&    2134.414&   2181.451&   2305.187&    2891.514&   2899.316&   2900.515 \\
  1801.301&    1898.196&    2134.508&   2181.461&   2305.198&    2891.525&   2899.327&   2900.526 \\
\vdots& \vdots& \vdots& \vdots& \vdots& \vdots& \vdots&  \vdots\\
\hline
& & & \multicolumn{2}{c}{direct2} & & &  \\
\hline
  1801.324&    1842.532&    1927.283&   2134.542&   2166.489&    2257.203&   2301.286&   2904.384 \\
  1802.297&    1842.543&    1931.200&   2134.553&   2166.500&    2257.214&   2301.297&   2905.345 \\
\vdots& \vdots& \vdots& \vdots& \vdots& \vdots& \vdots&  \vdots\\
\hline
& & & \multicolumn{2}{c}{direct3} & & & \\
\hline
  1801.337&    1829.289&    1925.290&   2131.554&   2151.492&    2259.242&   2305.232&   2908.536 \\
  1801.348&    1829.300&    1927.295&   2134.457&   2166.521&    2259.253&   2305.243&   2914.344 \\
\vdots& \vdots& \vdots& \vdots& \vdots& \vdots& \vdots&  \vdots\\
\hline
\hline
\end{tabular}

\medskip
Mid-exposure JDs are given as JD - 2450000, for each observed field
separately. Only a small sample of Table~\ref{table:images_jd} is
presented here. The entire table is available in the MNRAS electronic
issue.
\end{minipage}
\end{table*}

\subsection{Photometry}
\label{subsec:photometry}

Photometric processing was performed using IRAF packages and
routines\footnote{IRAF (Image Reduction and Analysis Facility) is
distributed by the National Optical Astronomy Observatories (NOAO),
which are operated by the Association of Universities for Research in
Astronomy (AURA), Inc., under cooperative agreement with the National
Science Foundation.}. Biasing and flat field correction were done with the
\texttt{CCDPROC} package with exposures taken nightly.

For object identification and astrometry we created for each field a
\emph{reference image} by combining the best 20 images.  A total of
\objnum objects were identified using the IRAF \texttt{daofind} task
\citep{Stetson87}. After X,Y coordinates were transformed to the
coordinate system of each of the images we applied aperture photometry
with the IRAF \texttt{phot} task in a fixed position mode.
Other photometric techniques we experimented with did not result in 
significantly different results.
Aperture radius was set to 3.5 pixels ($= 2.4$ arcsec) 
for all images, as this value is the typical PSF FWHM.

To remove systematic effects in our data we applied the newly
developed SysRem algorithm \citep{Tamuz05}, which can remove
systematic effects in large sets of photometric lightcurves. SysRem
succeeded to decrease the stellar scatter of the brightest objects by
up to about 50\%.

Astrometry was performed using the USNO A-2.0 star
catalog \citep{Monet98} and the reference image. CCD X,Y coordinates were
transformed to equatorial FK5 with second-order
polynomials. Residuals RMS were $\lesssim$ 0.45''
and no single coordinate residual exceeded 1.0''.

\begin{figure*}
\includegraphics[width=7.5cm,height=5.5cm]{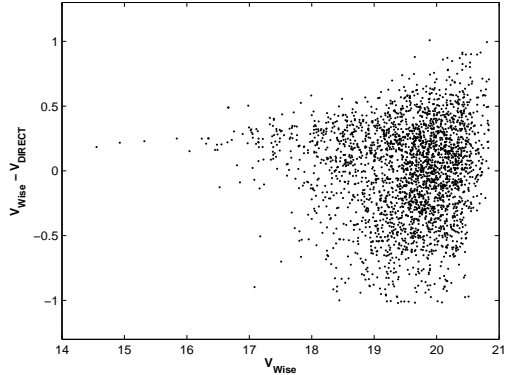}
\caption{Difference between Wise and DIRECT magnitudes vs. Wise magnitude 
for astrometrically matched objects.}
\label{fig:realmag}
\end{figure*}

Instrumental magnitude was transformed
to real magnitude using the catalog of \citet{Macri01a}.  This
catalog contains $V$ magnitudes of all objects observed by the DIRECT
first M33 observational campaign. A linear transformation from instrumental 
to real magnitude was derived using objects which were successfully
astrometrically matched with catalog objects, allowing a maximum
distance of 2''. Fig.~\ref{fig:realmag} shows the magnitude difference between
Wise and DIRECT magnitude vs. Wise derived magnitude for those objects.
The increased scatter in the magnitudes difference for faint stars results from 
the increased uncertainty in the magnitudes of those stars.

The RMS of the difference between Wise and DIRECT magnitudes is 0.3 -- 0.4 mag, 
varying slightly from field to field. This scatter is
similar to the scatter presented by \citet[][Figure 7, 2001b, Figure
6]{Mochejska01a} and found by \citet{Lee02}.

\subsection{Photometric Results and ON-Line Data}
\label{subsec:phot_res}

\begin{figure}
\centering
\includegraphics[height=10cm,width=12cm]{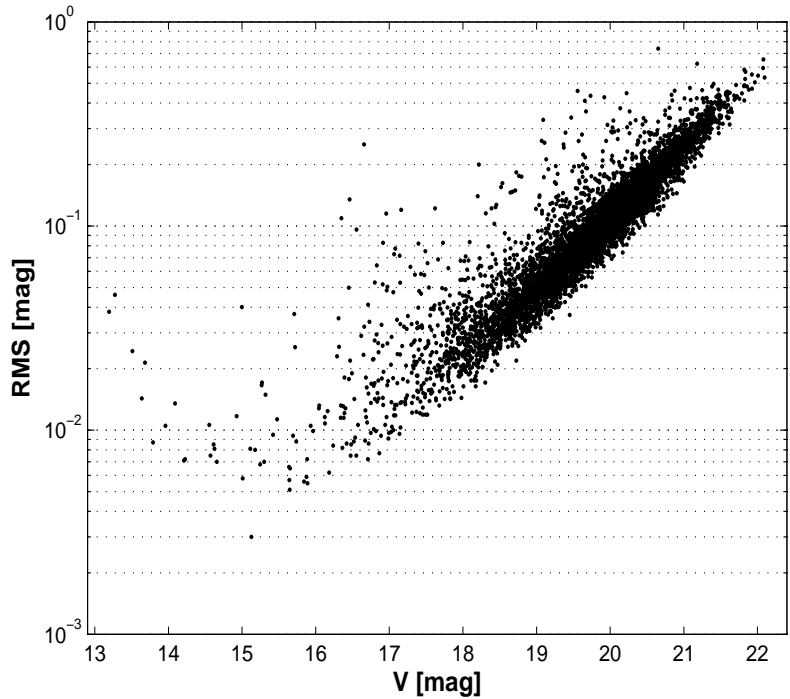}
\caption{ RMS against real $V$ magnitude for all \objnumf objects. The
decrease of the Poisson noise and read noise as objects become brighter, 
up to a
magnitude of $\sim\!16$, is the dominant feature of the diagram. For
brighter objects a residual systematic noise of $\sim\!7$
mmag is observed. The increased scatter for objects
brighter than $\sim\!14$th magnitude is due to CCD saturation.  }
\label{fig:rms_mag}
\end{figure}

Fig.~\ref{fig:rms_mag} presents the RMS vs. the averaged $V$ magnitude for all objects. 
Poisson and readout noises
are dominant for objects fainter than mag $\sim\!16$. For
brighter objects the scatter levels off and 
systematic noise at the level of $\sim \!7$ mmag becomes dominant.
For stars brighter
than $\sim\!14$ noise level rises again due to CCD saturation.  

Table~\ref{table:cat} lists the position, $V$ magnitude and RMS of all
\objnumf objects.
Lightcurves, consisting of a total of \allmeas individual
measurements are publicly available at the Wise
Observatory FTP server at {\tt wise-ftp.tau.ac.il:/pub/shporer/m33/}, 
which can be accessed through anonymous FTP. 
Each lightcurve is given as a separate text file,
named '{\tt lc}' with a 5-digit internal number extension. 
The listing of {\tt lc.12121} is given in Table \ref{table:lc}.

\begin{table*}
\begin{minipage}{75mm}
\caption{Object catalog}
\label{table:cat}
\begin{tabular}{ccccc}
\hline
Desig. &      R.A. &        Dec &       $V$ & $\sigma$\\
   &   hh mm ss.ss& dd mm ss.s & [mag]& [mag]\\
\hline
W30144& 01 32 55.66& 30 33 02.3& 20.63& 0.15\\
W31679& 01 32 55.68& 30 34 27.7& 18.79& 0.03\\
W31624& 01 32 55.69& 30 35 34.8& 17.61& 0.04\\
W30656& 01 32 55.73& 30 31 41.4& 19.65& 0.08\\
W31717& 01 32 55.73& 30 34 30.1& 18.48& 0.03\\
W30559& 01 32 55.73& 30 39 31.3& 20.08& 0.12\\
W31082& 01 32 55.74& 30 39 22.4& 20.36& 0.17\\
W31078& 01 32 55.77& 30 39 12.7& 19.71& 0.08\\
W31484& 01 32 55.81& 30 37 13.0& 16.42& 0.01\\
W30901& 01 32 55.83& 30 35 24.4& 19.80& 0.09\\
\vdots& \vdots& \vdots& \vdots&  \vdots\\
\hline 
\end{tabular}

\medskip
Columns contain (1) designation, (2) R.A. (J2000.0), (3) Dec (J2000.0),
(4) $V$ magnitude and (5) magnitude RMS. Only a small sample of
Table~\ref{table:cat} is presented here. The entire table is available
in the MNRAS electronic issue.
\end{minipage}
\end{table*}

\begin{table*}
\begin{minipage}{44mm}
\caption{Example of an on-line lightcurve text file.}
\label{table:lc}
\begin{tabular}{ccc}
\hline
1809.4479&  20.511&  0.146\\
1809.4589&  20.591&  0.158\\
1813.5006&  20.358&  0.131\\
1819.2919&  20.490&  0.169\\
1819.3029&  20.697&  0.184\\
1819.3735&  20.522&  0.113\\
1819.3844&  20.493&  0.120\\
1820.2384&  20.437&  0.243\\
1820.2494&  20.267&  0.170\\
1820.3184&  20.595&  0.154\\
\vdots&     \vdots& \vdots\\
\hline
\end{tabular}

\medskip
Columns contain (1) mid-exposure JD - 2450000, (2) $V$ magnitude and (3) 
magnitude uncertainty. Only part of the lightcurve is presented here,
for guidance regarding form and content of all lightcurve text files. 
 
\end{minipage}
\end{table*}

\section{Search for Variability}
\label{sec:temp_anal} 

Each Lightcurve was searched first for periodic modulation, and then for
non-periodic variability. For stars with available DIRECT data, we also
applied periodic analysis to the combined data.

\subsection{Periodicity Detection}
\label{subsec:per_detect}

Periodicity search was applied to all lightcurves with the Analysis
of Variance (AoV) algorithm of \citet{Czerny89}. For each trial
frequency, $\nu$, data was folded with the corresponding period and
then binned into 10 bins, using equal as possible number of points per
bin. Two variances were calculated: Binned lightcurve variance, $s_1^2(\nu)$, 
and the sum of bins internal variance, $s_2^2(\nu)$. Periodogram value, 
$S(\nu)$, was taken as the ratio of those variances, $s_1^2(\nu)/s_2^2(\nu)$.

For each lightcurve, we defined \SIG\ as the value of the highest periodogram 
peak divided by periodogram average:

\begin{equation}
\mathcal{S} = \frac{max_{\nu}S(\nu)}{<S(\nu)>_{\nu}}\,.
\end{equation}
We consider the value of \SIG\ as an indicator of the significance of the
detection of a periodicity within the lightcurve.

To estimate the significance of the periodicity detection 
we computed \SIG\ for 100 random permutations of
every lightcurve in each of the three fields, obtaining an \SIG\
distribution consisting of $\sim 2 \times 10^5$ elements per field. 
We defined \HS\ to be the
percentage of randomly permuted lightcurves with \emph{higher} \SIG\
values. Stars with \HS\ smaller than $0.01\%$ were flagged as
periodic. 
This threshold gives an expectation of one false detection
for every $10^4$ lightcurves, or, $0.64$ for our entire sample.

A total of \allpernum periodic variables were detected.
The period uncertainty, 
${\Delta}P$, was defined as the FWHM of the periodogram peak.
In order to derive the amplitude we applied an iterative running-median 
to the phased lightcurve and took half the magnitude difference between 
maximum and minimum brightness. 

\begin{figure}
\includegraphics[height=7in,width=6.5in]{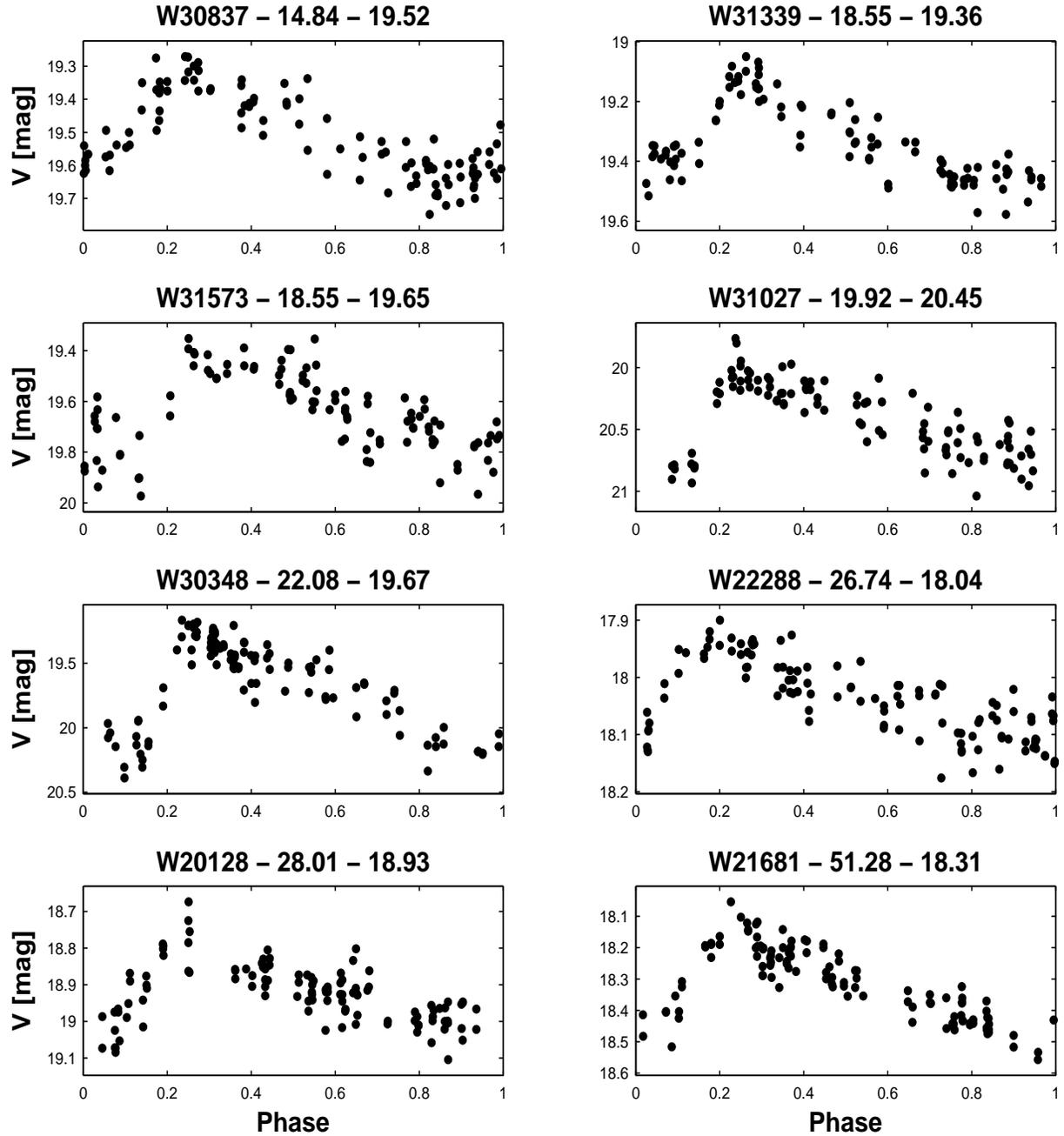}
\caption{Lightcurves of the \cephnew new Cepheids detected here. Magnitude
is plotted against phase. Title of each lightcurve is composed of 
(from left to right) designation, period, in days, and mean $V$
magnitude.}
\label{fig:cephlc}
\end{figure}

We have classified \cephnum periodic variables as Cepheids
by examining the period, amplitude and shape of all periodic variables.
In particular, we examined the relative duration of increasing and decreasing 
brightness, in order to detect stars with an increasing phase substantially 
shorter than the decreasing phase. 
Comparing with the publicly 
available DIRECT catalogs and the SIMBAD astronomical database we find that 
\cephnew out of the \cephnum Cepheids are new, and \knownceph are previously 
known Cepheids.
The \cephnew new Cepheids are listed in Table~\ref{table:newceph} and 
plotted in Fig.~\ref{fig:cephlc}. 
Five of the new Cepheids were not identified before as variables, and the 
other \cephnewclass were classified as variables and had no known period.
The \cephnewvar new Cepheids are positioned in our direct3 field and in  
the DIRECT M33C field. Variables of this field were not reported 
by the DIRECT project, although Table~9 of 
\citet[reporting variables in fields M33A and M33B]{Macri01b}
includes 15 M33C variables. (The \knownceph known Cepheids are included 
in Table~\ref{table:complist} which lists all variables identified here.)  

\begin{table*}
\begin{centering}
\begin{minipage}{120mm}
\caption{\label{table:newceph} List of \cephnew new Cepheids. Five of the new 
Cepheids were not identified before as variables and the other 3 
were classified as variables and had no known period.}
\begin{tabular}{lcccccllc}
\hline
\multicolumn{1}{c}{Desig.}& \multicolumn{1}{c}{R.A.}& 
\multicolumn{1}{c}{Dec.}& \multicolumn{1}{c}{$V$}& 
\multicolumn{1}{c}{$\sigma$}& \multicolumn{1}{c}{Amp.}& 
\multicolumn{1}{c}{Period}& \multicolumn{1}{c}{${\Delta}P$}& 
\multicolumn{1}{c}{Comments}\\

& \multicolumn{1}{c}{hh mm ss.ss}& 
\multicolumn{1}{c}{dd mm ss.s}& \multicolumn{1}{c}{[mag]}&
\multicolumn{1}{c}{[mag]}& \multicolumn{1}{c}{[mag]}& 
\multicolumn{1}{c}{[days]} & \multicolumn{1}{c}{[days]} & \\

\hline
\hline
\multicolumn{9}{c}{New Variables}  \\
\hline
W30837&  1 33 27.66&  30 34 24.0&   19.52&  0.12&  0.15&   14.837&   0.073& 1\\
W31339&  1 32 57.55&  30 38 46.5&   19.36&  0.15&  0.16&   18.553&   0.096& 1\\
W31573&  1 33 16.45&  30 36 58.2&   19.65&  0.15&  0.18&   18.553&   0.091& 1\\
W31027&  1 33 30.15&  30 38 04.8&   20.45&  0.34&  0.35&   19.920&   0.100& 1\\
W30348&  1 33 30.40&  30 35 55.8&   19.67&  0.36&  0.45&    22.08&    0.13& 1\\
\hline
\multicolumn{9}{c}{New Periodic Variables}  \\
\hline
W22288&  1 33 57.56&  30 38 44.8&   18.04&  0.07&  0.08&    26.74&    0.16&\\
W20128&  1 34 15.47&  30 31 06.7&   18.93&  0.08&  0.10&    28.01&    0.17&\\
W21681&  1 33 39.92&  30 35 08.2&   18.31&  0.12&  0.15&    51.28&    0.91&\\
\hline
\end{tabular}

\medskip
Columns contain (1) Designation, (2) J2000.0 R.A., (3) J2000.0 Dec., 
(4) mean $V$ magnitude, (5) magnitude RMS, (6) amplitude, 
(7) period, (8) period uncertainty and (9) comments. \\
Comments: 1: Position in DIRECT M33C field.

\end{minipage}
\end{centering}
\end{table*}

\begin{figure}
\includegraphics[height=5.25in,width=6.5in]{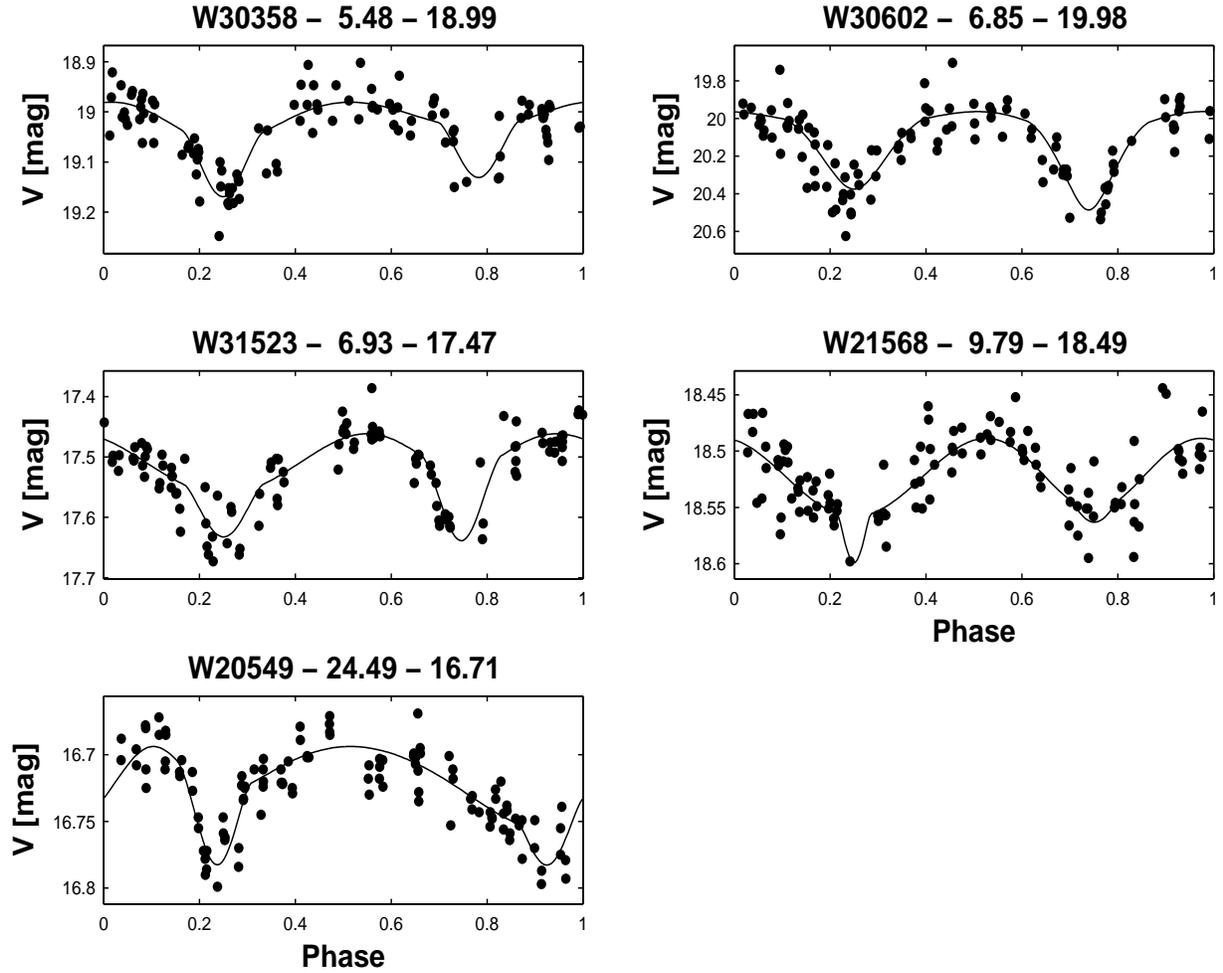}
\caption{Lightcurves of the \ebnew new EBs detected here.
Magnitude is plotted against phase. EBAS lightcurve solution is presented 
by a solid line. Title of each lightcurve is composed of (from left to right) 
Designation, EBAS period, in days, and EBAS out-of-eclipse $V$ 
magnitude.}
\label{fig:eblc}
\end{figure}

In order to detect EBs we applied the Eclipsing Binary
Automated Solver (\citealt*{Tamuz06}, \citealt*{Mazeh06}, hereafter
EBAS) to all periodic variables that were not identified as Cepheids. 
EBAS has its own built in goodness-of-fit estimator, 
the \emph{alarm} --- \A, which we used to identify the true EBs. 
The advantage of the alarm over the classical $\chi^2$, as a 
goodness-of-fit estimator, results from accounting for measurements 
\emph{order}, by considering maximal series of \emph{consecutive}, 
mean-subtracted measurements with the same sign. 
We manually inspected
all objects with \A~$ < 1.0$, in order to remove spurious EB classifications. 
We identified \ebnum systems as EBs, 
of which \ebnew are new, presented in Fig.~\ref{fig:eblc} and listed in 
Table~\ref{table:neweb}. Those \ebnew include
four new variables and one 
new classification of a known, unclassified periodic
object (W21568). Table~\ref{table:ebas} lists the EBAS solution
parameters of all \ebnum EBs detected here.
The \knowneb already known EBs are included in 
Table~\ref{table:complist}.

\begin{table*}
\begin{centering}
\begin{minipage}{80mm}
\caption{\label{table:neweb} List of \ebnew new eclipsing-binaries. 
Three of those were not identified before as variables, and the other one 
was identified as an unclassified periodic.}
\begin{tabular}{lccccc}
\hline
\multicolumn{1}{c}{Desig.}& \multicolumn{1}{c}{R.A.}& 
\multicolumn{1}{c}{Dec.}&\multicolumn{1}{c}{$V$}& 
\multicolumn{1}{c}{$\sigma$}& \multicolumn{1}{c}{Comments}  \\
& \multicolumn{1}{c}{hh mm ss.ss}& \multicolumn{1}{c}{dd mm ss.s}& 
\multicolumn{1}{c}{[mag]}&\multicolumn{1}{c}{[mag]} &\\

\hline
\hline
\multicolumn{6}{c}{New Variables}  \\
\hline
W30358&  1 32 57.31&  30 36 07.6&   19.04&  0.07& 1\\
W31523&  1 33 29.88&  30 31 47.5&   17.52&  0.07& 1\\
W30602&  1 33 35.31&  30 40 24.7&   20.15&  0.19& 1\\
W20549&  1 33 58.67&  30 35 26.6&   16.73&  0.03& \\
\hline
\multicolumn{6}{c}{New Classification}  \\
\hline
W21568&  1 33 54.80&  30 32 49.0&   18.52&  0.04& \\
\hline
\end{tabular}

\medskip 
Columns contain (1) Designation, (2) J2000.0 R.A., (3) J2000.0 Dec., 
(4) mean $V$ magnitude, (5) magnitude RMS and (6) comments.\\
Comments: 1: Position in DIRECT M33C field.

\end{minipage}
\end{centering}
\end{table*}

\begin{table*}
\begin{centering}
\begin{minipage}{165mm}
\caption{\label{table:ebas} EBAS orbital solutions for the \ebnum 
eclipsing-binaries identified here.}
\begin{tabular}{rrrrrrrrrrrr}
\hline
\multicolumn{1}{c}{Desig.}& \multicolumn{1}{c}{\A}& 
\multicolumn{1}{c}{$V$}& \multicolumn{1}{c}{Period}& 
\multicolumn{1}{c}{$r_t$}& \multicolumn{1}{c}{$k$}& 
\multicolumn{1}{c}{$J_s$}& \multicolumn{1}{c}{$x$}& 
\multicolumn{1}{c}{$e\,cos\,w$}& \multicolumn{1}{c}{$e\,sin\,w$}& 
\multicolumn{1}{c}{$A_p$}& \multicolumn{1}{c}{$A_s$}\\  
& & \multicolumn{1}{c}{[mag]}& \multicolumn{1}{c}{[days]}& & & & & & & &\\  
\hline
W21612& -0.09&  19.4681&     2.337324&       0.706&       0.2&           1.9&           0.79&         0.014&        -0.018&       0.995&       1.0000 \\
        & & $\pm$3.2e-02& $\pm$8.5e-05& $\pm$9.7e-02& $\pm$1.2e+00& $\pm$2.6e+00& $\pm$1.3e-01& $\pm$3.1e-02& $\pm$4.4e-02& $\pm$2.6e-02& $\pm$2.3e-03 \\
W11486& 0.12&  18.359&       2.70811&      0.81&         0.93&          0.75&         0.800&        0.034&        -0.037&       0.99&        1.000 \\
      & & $\pm$3.1e-02& $\pm$1.1e-04& $\pm$1.0e-01& $\pm$5.8e-01& $\pm$2.7e-01& $\pm$8.1e-02& $\pm$2.4e-02& $\pm$4.5e-02& $\pm$2.1e-01& $\pm$1.4e-02 \\
W11491& -0.22&  18.980&       3.84684&         0.640&       1.44&         1.27&         0.64&        0.002&          0.08&         1.0000&      1.00 \\
      & & $\pm$7.2e-02& $\pm$3.2e-04& $\pm$9.7e-02& $\pm$8.0e-01& $\pm$8.4e-01& $\pm$1.1e-01& $\pm$4.7e-02& $\pm$1.0e-01& $\pm$3.9e-03& $\pm$1.7e-01 \\
W10764& 0.55&  20.318&      4.432920&       0.731&        0.59&        1.43&         -0.09&        -0.060&        0.05&         0.016&   0.08 \\
      & & $\pm$7.0e-02& $\pm$7.1e-05& $\pm$5.7e-02& $\pm$5.6e-01& $\pm$6.0e-01& $\pm$1.3e-01& $\pm$2.3e-02& $\pm$1.0e-01& $\pm$1.3e-02& $\pm$1.6e-01 \\
W20942& 0.53&  20.739&      5.09566&        0.670&       1.06&         0.67&        -0.001&        -0.015&      -0.06&        1.0000&   0.34 \\
      & & $\pm$4.9e-02& $\pm$1.6e-04& $\pm$6.8e-02& $\pm$2.3e-01& $\pm$1.9e-01& $\pm$8.5e-02& $\pm$2.5e-02& $\pm$1.2e-01& $\pm$9.2e-03& $\pm$2.6e-01 \\
W30358& 0.16&  18.990&      5.48325&        0.634&       0.68&         1.36&         0.661&         0.047&      0.061&         0.000&   1.000 \\
      & & $\pm$3.7e-02& $\pm$3.3e-04& $\pm$3.7e-02& $\pm$4.4e-01& $\pm$4.7e-01& $\pm$5.2e-02& $\pm$1.2e-02& $\pm$5.6e-02& $\pm$3.9e-02& $\pm$1.6e-03 \\
W30051& 0.78&  18.5282&     6.624649&       0.6751&      2.64&         0.67&        -0.015&        -0.0645&      0.024&         0.07&   1.00 \\
      & & $\pm$7.3e-03& $\pm$1.1e-05& $\pm$9.7e-03& $\pm$9.0e-01& $\pm$1.2e-01& $\pm$5.1e-02& $\pm$5.4e-03& $\pm$1.4e-02& $\pm$2.0e-01& $\pm$1.4e-01 \\
W30602& -0.16&  19.980&      6.85114&      0.785&        0.93&         0.99&	     0.41&         -0.017&      -0.106&          1.00&   0.00 \\
      & & $\pm$6.6e-02& $\pm$4.1e-04& $\pm$6.1e-02& $\pm$5.1e-01& $\pm$1.6e-01& $\pm$1.0e-01& $\pm$1.7e-02& $\pm$9.7e-02& $\pm$2.8e-01& $\pm$2.6e-01 \\
W31523& -0.18&  17.473&      6.92557&       0.634&       1.71&         0.73&         0.784&        -0.0056&      -0.106&        0.005&   1.000 \\
      & & $\pm$1.6e-02& $\pm$1.2e-04& $\pm$3.3e-02& $\pm$4.0e-01& $\pm$6.6e-02& $\pm$1.7e-02& $\pm$5.4e-03& $\pm$1.6e-02& $\pm$3.4e-02& $\pm$5.7e-02 \\
W22214& -0.23&  18.582&     8.77401&        0.455&       3.8&         0.89&         0.32&          -0.031&      -0.001&         0.902&   0.99983 \\
      & & $\pm$2.4e-02& $\pm$4.1e-04& $\pm$1.5e-02& $\pm$1.3e+00& $\pm$3.7e-01& $\pm$1.7e-01& $\pm$1.0e-02& $\pm$3.9e-02& $\pm$2.6e-02& $\pm$1.5e-04 \\
W21568& -0.20&  18.493&      9.7870&        0.549&        4.0&         0.050&       0.947&          0.002&      -0.067&          1.000&   0.13 \\
      & & $\pm$2.9e-02& $\pm$1.3e-03& $\pm$3.9e-02& $\pm$1.1e+00& $\pm$7.2e-02& $\pm$5.4e-02& $\pm$3.1e-02& $\pm$3.7e-02& $\pm$3.3e-02& $\pm$2.2e-01 \\
W20549&  0.27&  16.71&        24.4920 &     0.652&         1.16 &       1.343 &       0.7938 &       0.2464 &     0.0600 &       0.940 &   0.36  \\  
      & & $\pm$1.1e-1& $\pm$3.9e-3& $\pm$2.1e-2& $\pm$1.1e-1& $\pm$5.2e-2& $\pm$9.2e-3& $\pm$6.7e-3& $\pm$8.1e-3 & $\pm$6.1e-2 & $\pm$1.0e-1 \\ 

\hline
\end{tabular}

\medskip
Columns contain (1) Designation, (2) alarm, (3) out-of-eclipse $V$ magnitude, 
(4) period, in days, (5) relative sum of radii, (6) radii ratio, (7) surface brightness ratio, 
(8) impact parameter, (9) eccentricity, $e$, multiplied by cos$w$ and (10) by sin$w$, 
where $w$ is the periastron longitude, (11) bolometric reflection coefficient for the 
primary binary component and (12) for the secondary component.
\end{minipage}
\end{centering}
\end{table*}

\begin{figure}
\includegraphics[height=7in,width=6.5in]{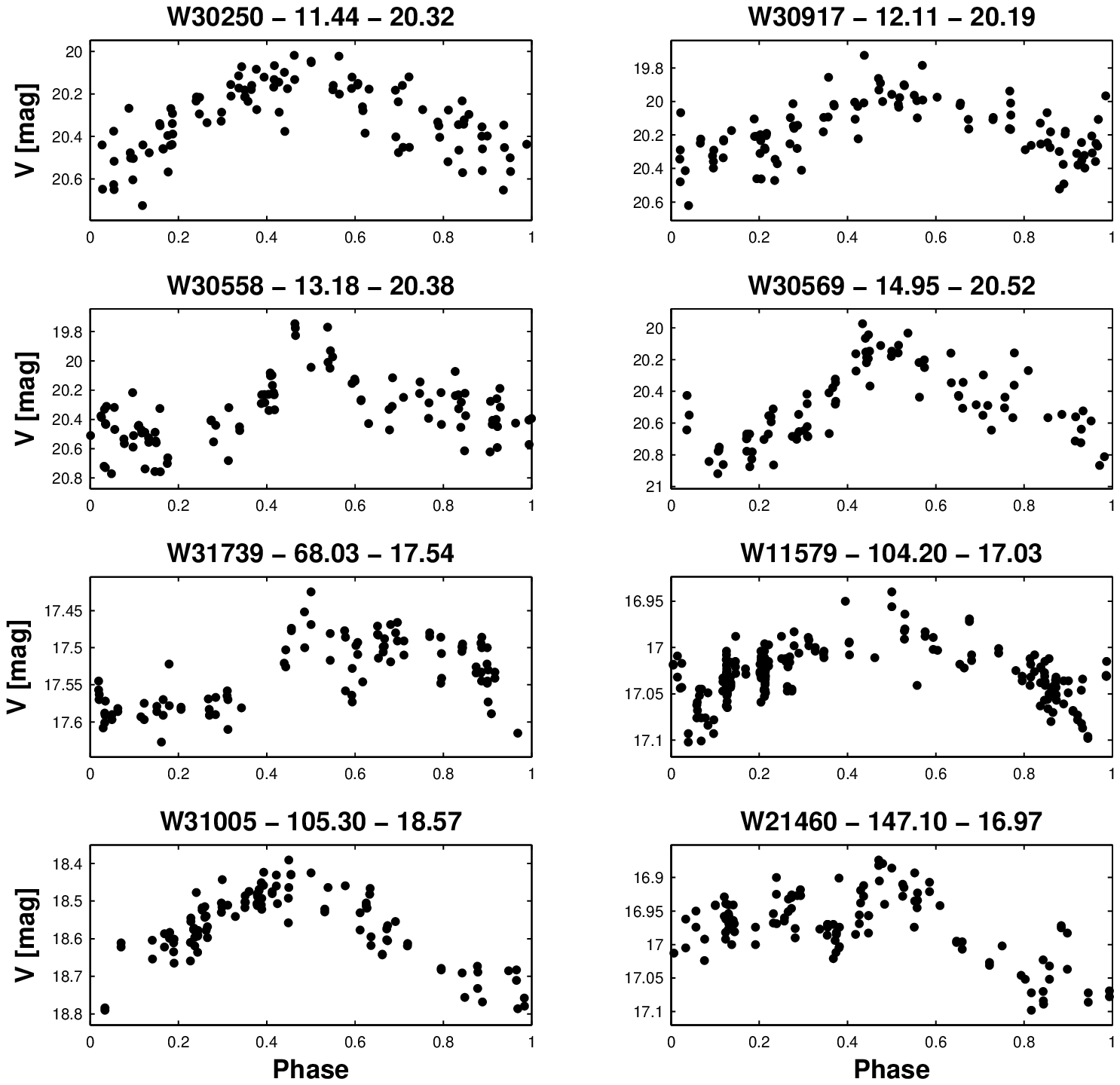}

\caption{Lightcurves of eight new unclassified periodics, out of 
the \pernew new periodic variables identified here.
Magnitude is plotted against phase. Title of each lightcurve is composed 
of (from left to right) designation, period, in days, and mean 
$V$ magnitude.}
\label{fig:perlc}
\end{figure}

Periodic variables not identified as Cepheids or EBs were
noted as \emph{unclassified periodic}. Those include \pernum objects,
\pernew of which were not known before as periodic variables, 
listed on Table~\ref{table:newper}. (The \knownper
known periodic variables are included in Table~\ref{table:complist}.) 
The \pernew new unclassified periodics include \pernewvar new variables 
(not known before as variables) and \pernewclass previously known as 
non-periodic variables. 
A sample of 8 new unclassified periodic variables is presented in 
Fig.~\ref{fig:perlc}. 

In addition, we have determined an improved period, of 175.4 days, for W11984 
where the previously known period was 458 days \citep{Kinman87}.

\begin{table*}
\begin{centering}
\begin{minipage}{120mm}
\caption{\label{table:newper} List of \pernew new unclassified periodic 
variables. Twenty of those were not identified before as variables. The 
other 8 are known variables without a known period.}
\begin{tabular}{lccccc r@{.}l r@{.}l c}
\hline
\multicolumn{1}{c}{Desig.}& \multicolumn{1}{c}{R.A.}& 
\multicolumn{1}{c}{Dec.}&\multicolumn{1}{c}{$V$}& 
\multicolumn{1}{c}{$\sigma$}& \multicolumn{1}{c}{Amp.}& 
\multicolumn{2}{c}{\ \ \ Period}& \multicolumn{2}{c}{ ${\Delta}P$ }&
\multicolumn{1}{c}{Comments}\\
& \multicolumn{1}{c}{hh mm ss.ss}& 
\multicolumn{1}{c}{dd mm ss.s}& \multicolumn{1}{c}{[mag]}&
\multicolumn{1}{c}{[mag]}& \multicolumn{1}{c}{[mag]}& 
\multicolumn{2}{c}{\ \ \ [days]}& \multicolumn{2}{c}{[days] }& \\
\hline
\hline
\multicolumn{11}{c}{New Variables}  \\
\hline
W10109&  1 34 15.57&  30 41 10.1&   18.61&  0.04&   0.04&   9&718&   0&021& \\
W30344&  1 33 27.37&  30 35 51.5&   20.81&  0.24&   0.25&  11&211&   0&021& 1\\
W30250&  1 33 25.58&  30 34 27.0&   20.32&  0.18&   0.19&  11&442&   0&047& 1\\
W30917&  1 33 18.48&  30 35 34.3&   20.19&  0.18&   0.18&  12&107&   0&061& 1\\
W30558&  1 33 31.65&  30 39 32.0&   20.38&  0.26&   0.27&  13&175&   0&048& 1\\
W30569&  1 33 32.17&  30 39 46.2&   20.52&  0.29&   0.33&  14&948&   0&046& 1\\
W20380&  1 34 16.05&  30 33 44.9&   16.79&  0.02&   0.03&  15&773&   0&063& \\
W12052&  1 34 00.90&  30 40 25.1&   18.25&  0.04&   0.05&  26&46&    0&34& \\
W12073&  1 33 45.15&  30 44 19.0&   18.02&  0.04&   0.04&  52&9&     1&2& \\
W31739&  1 33 12.34&  30 38 48.7&   17.54&  0.05&   0.05&  68&03&    0&80& 1\\
W31230&  1 33 11.16&  30 34 21.9&   16.80&  0.03&   0.04&  99&3&     4&4& \\
W11579&  1 33 48.32&  30 42 41.4&   17.03&  0.03&   0.03& 104&2&     3&1& \\
W31005&  1 33 05.77&  30 37 20.3&   18.57&  0.09&   0.15& 105&3&     3&4& 1\\
W31047&  1 33 15.76&  30 38 22.1&   17.96&  0.05&   0.07& 117&6&     3&8& 1\\
W30803&  1 33 00.09&  30 33 41.4&   19.27&  0.10&   0.09& 140&8&     4&8& 1\\
W21460&  1 33 50.96&  30 38 19.4&   16.97&  0.05&   0.08& 147&1&     5&0& \\
W30292&  1 32 59.34&  30 35 05.1&   17.91&  0.04&   0.04& \multicolumn{2}{l}{222 \ \ \ \ }& \multicolumn{2}{l}{$\!\!\!$13 }& 1\\
W11898&  1 34 10.91&  30 41 41.6&   18.70&  0.05&   0.06& \multicolumn{2}{l}{270 \ \ \ \ }& \multicolumn{2}{l}{$\!\!\!$34 }& \\
W31416&  1 33 12.60&  30 32 52.5&   19.51&  0.20&   0.31& \multicolumn{2}{l}{278 \ \ \ \ }& \multicolumn{2}{l}{$\!\!\!$29 }& 1\\
\hline
\multicolumn{11}{c}{New Periodic Variables}  \\
\hline
W30043&  1 33 36.60&  30 31 55.0&   19.36&  0.09&   0.09&  13&624&   0&068& \\
W21058&  1 33 41.34&  30 32 12.9&   18.24&  0.06&   0.06&   20&12&    0&16& \\
W21193&  1 33 58.33&  30 34 29.7&   18.04&  0.04&   0.04&   36&50&    0&29& \\
W21387&  1 34 05.42&  30 37 19.7&   19.16&  0.10&   0.13&   135&1&     6&5& \\
W20509&  1 33 55.58&  30 35 01.0&   17.39&  0.04&   0.06&   144&9&     7&0& \\
W11870&  1 33 52.39&  30 39 08.4&   16.55&  0.10&   0.15& \multicolumn{2}{l}{263\ \ \ \ }&\multicolumn{2}{l}{$\!\!\!$10 }& \\
W10678&  1 33 58.93&  30 41 38.5&   17.11&  0.02&   0.04& \multicolumn{2}{l}{286\ \ \ \ }& \multicolumn{2}{l}{$\!\!\!$26 }& \\

\hline
\end{tabular}

\medskip
Columns contain (1) Designation, (2) J2000.0 R.A., (3) J2000.0 Dec., 
(4) mean $V$ magnitude, (5) magnitude RMS,
(6) period, (7) period uncertainty ,(8) amplitude and (9) comments.\\
Comments: 1: Position in DIRECT M33C field. 
\end{minipage}
\end{centering}
\end{table*}

\subsection{Non-Periodic Variability Detection}
\label{subsec:var_detect}

All non-periodic lightcurves were searched for variability using the 
\emph{alarm} statistic, \A, 
introduced by \cite{Tamuz06}. Here, the alarm is used to detect 
variable lightcurves 
by estimating the goodness-of-fit of a constant function.
The significance of the detection of the variability was estimated 
by a permutation test, as was done for the periodicity detection.
For each single lightcurve, we generated $10^5$ random permutations and 
calculated their alarm value. For each real lightcurve we defined \HA\ to 
be the percentage of randomly permuted lightcurves with \emph{higher} \A\ 
values. Lightcurves with \HA\ smaller than $0.01\%$ were flagged as 
non-periodic variables. 
 
\begin{figure}
\includegraphics[height=7in,width=6.5in]{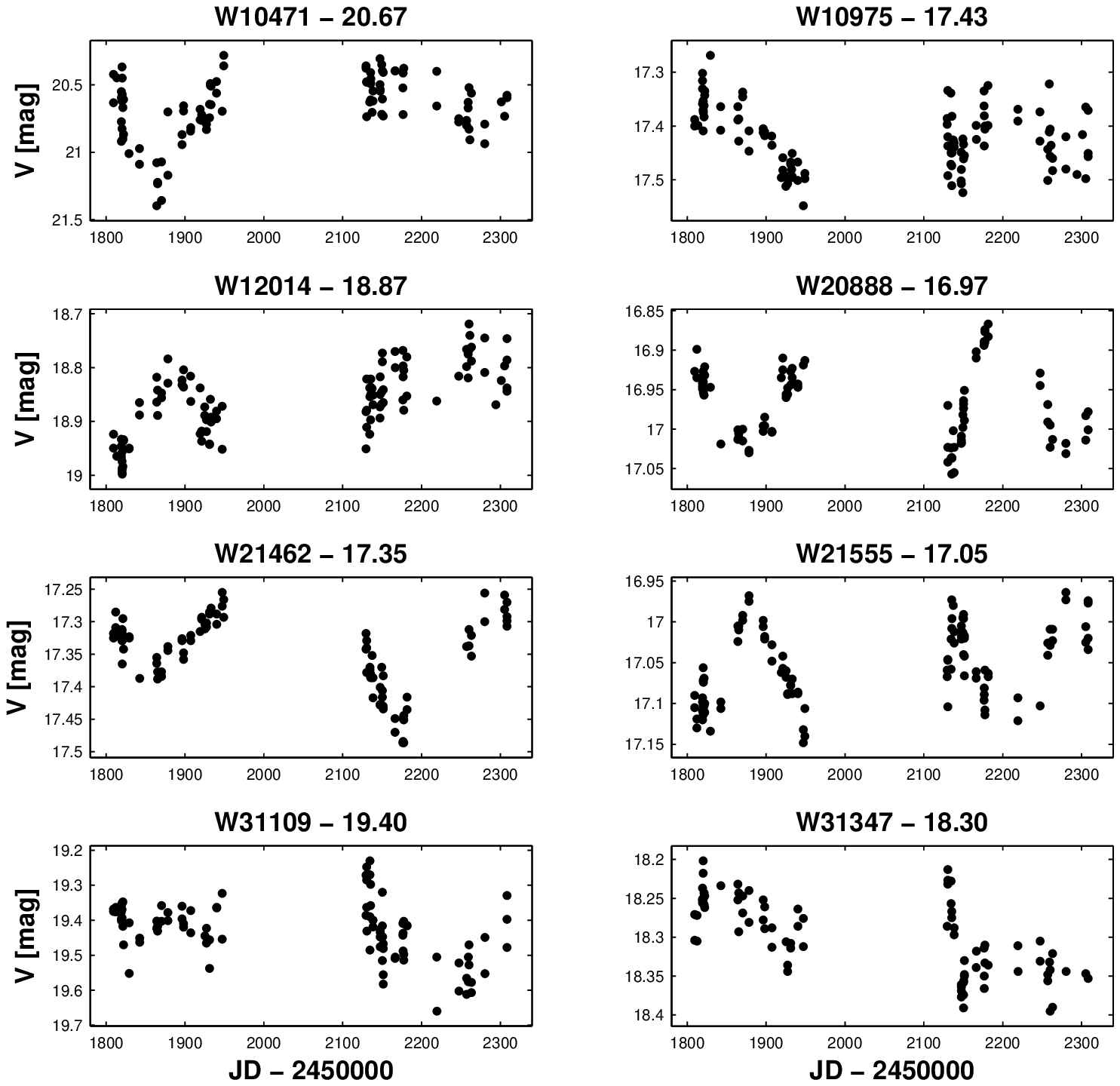}
\caption{Lightcurves of eight new non periodic variables, out of the 
\nonpernew new non-periodics detected here. Magnitude is plotted vs. JD - 2450000. 
Title of each lightcurve is composed of (from left to right)
designation and mean $V$ magnitude. Data of only the first two observational 
seasons is presented on the above plots, to make the variability more apparent.}
\label{fig:varlc}
\end{figure}

A total of \nonpernum non-periodic variables were detected. 
Those include \nonpernew new variables, listed on Table~\ref{table:newvar}.
(The \knownnonper previously known are included in Table~\ref{table:complist}.) 
A sample of eight new non-periodic variables is presented on 
Fig.~\ref{fig:varlc}. 

\begin{table*}
\begin{minipage}{90mm}
\caption{List of \nonpernew new non-periodic variables.}
\label{table:newvar}

\begin{tabular}{cccccc}
\hline
\multicolumn{1}{c}{Desig.}& \multicolumn{1}{c}{R.A.}& 
\multicolumn{1}{c}{Dec.}& \multicolumn{1}{c}{$V$}& 
\multicolumn{1}{c}{$\sigma$}& \multicolumn{1}{c}{Comments}\\
& \multicolumn{1}{c}{hh mm ss.ss}& \multicolumn{1}{c}{dd mm ss.s}& 
\multicolumn{1}{c}{[mag]}& \multicolumn{1}{c}{[mag]}&\\

\hline
W31624&  1 32 55.69&  30 35 34.8&   17.61&  0.04&  1\\
W31074&  1 32 56.73&  30 39 04.0&   19.22&  0.12&  1\\
W31279&  1 32 57.86&  30 35 55.1&   19.13&  0.10&  1\\
W30354&  1 32 58.19&  30 36 06.4&   18.80&  0.17&  1\\
W31223&  1 32 59.58&  30 34 05.8&   19.23&  0.08&  1\\
W30648&  1 32 59.70&  30 31 37.0&   19.84&  0.12&  1\\
W31070&  1 32 59.74&  30 38 55.1&   18.06&  0.06&  1\\
W30214&  1 33 00.94&  30 34 04.0&   19.19&  0.08&  1\\
W30877&  1 33 01.02&  30 35 00.7&   18.79&  0.08&  1\\
W30185&  1 33 01.70&  30 33 29.6&   19.55&  0.10&  1\\
\vdots&   \vdots&        \vdots&       \vdots&  \vdots&  \vdots \\    
\hline

\end{tabular}

\medskip
Columns contain (1) Designation, (2) J2000.0 R.A., (3) J2000.0 Dec., 
(4) mean $V$ magnitude, (5) magnitude RMS and (6) Comments.\\ 
Comments: 1: Position in DIRECT M33C field. 2: Position outside 
all three DIRECT fields.\\ 
Only a small sample of Table~\ref{table:newvar} is presented here. 
The entire table is available in the MNRAS electronic issue.  
\end{minipage}
\end{table*}

\begin{figure}
\centering
\includegraphics[height=9cm,width=10cm]{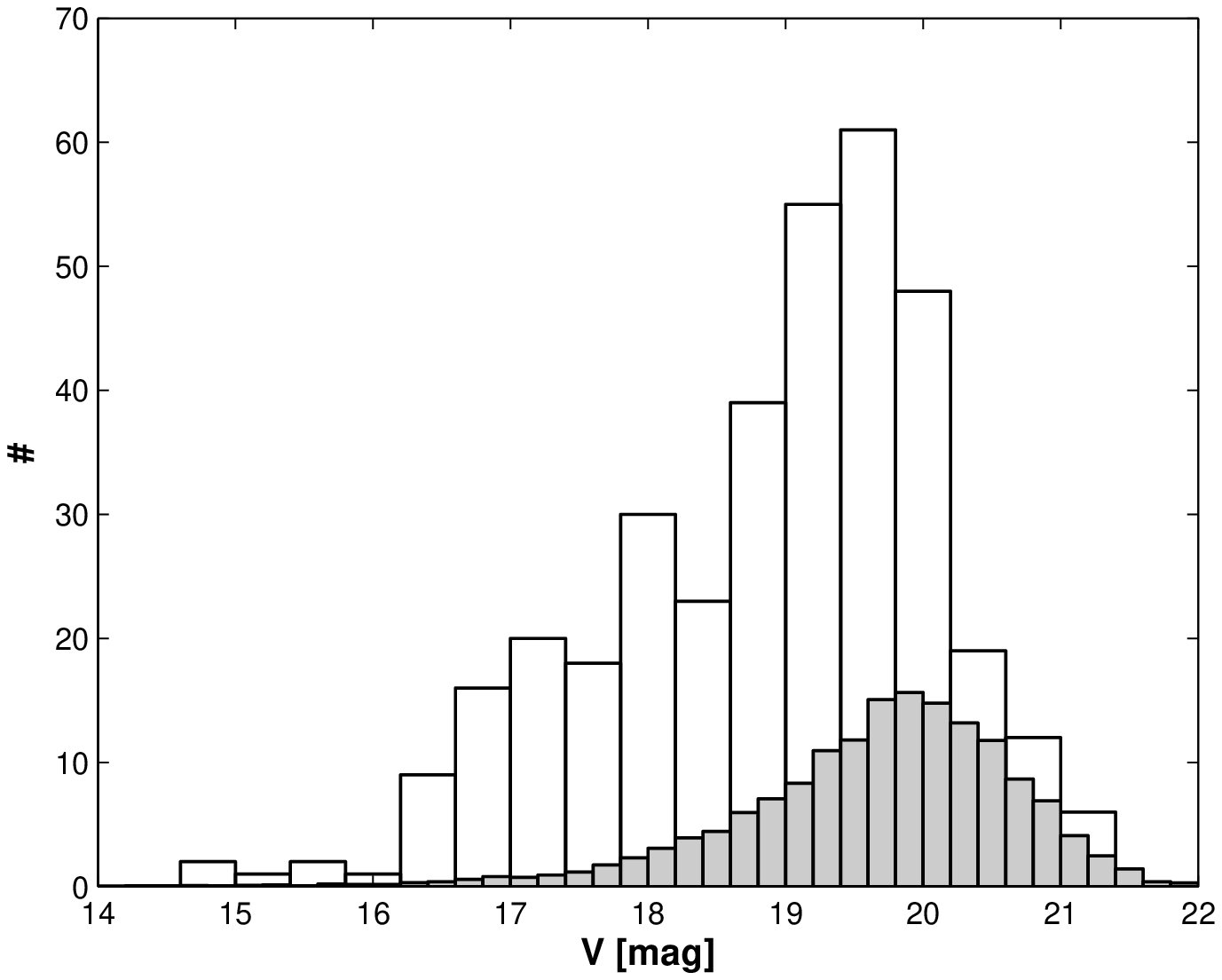}
\caption{Magnitude histogram of all variables detected here, showing a 
completeness magnitude of about 19.5. Corresponding (scaled) histogram for 
\emph{all} objects is presented in gray.}
\label{fig:maghist} 
\end{figure}

Fig.~\ref{fig:maghist} presents a mean $V$ magnitude histogram of all variable 
(periodic and non-periodic) objects detected here, indicating a completeness 
magnitude of about 19.5. For comparison, the corresponding histogram for all 
objects, where bin hight was reduced by a factor of 40, is also presented, in 
gray.

\subsection{Combined Lightcurves}
\label{subsec:comb_lc} 

We combined the data presented in this paper with the available
data for DIRECT variables from \cite{Macri01a} and \cite{Mochejska01a,Mochejska01b}.
Data from all three sources were combined to create 
$\la$~7 years time span lightcurves, although with large gaps, 
of up to 500 $V$ measurements each. 
Combining data from several
telescopes requires fine calibration, removing possible 
zero point differences, 
when searching for variables in particular.

To derive a periodogram we fitted the data from the three telescopes
with a few harmonics for each frequency together with different zero 
points for the three data sources. Those zero points were fitted separately 
for each variable. Since there is a large scatter in the magnitude difference
of the same stars in different telescopes (See Fig.~\ref{fig:realmag} and 
\citet[][Figure 7, 2001b, Figure 6]{Mochejska01a}) taking the same zero point 
shift for all lightcurves would not be accurate enough.
  
\begin{figure}
\centering
\includegraphics[height=14cm,width=10cm]{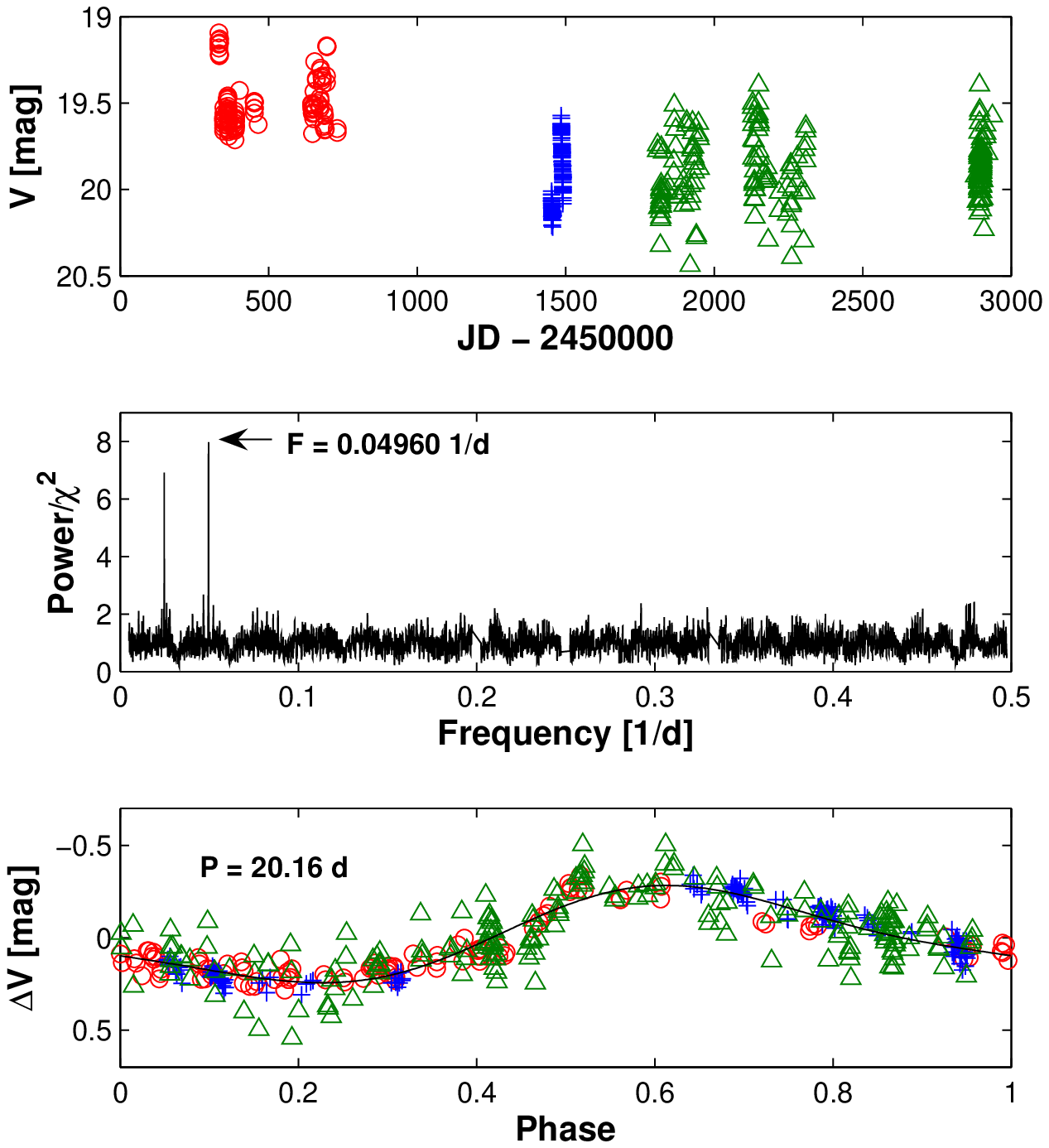}\\
\caption{Combined lightcurve analysis of W10821 and its DIRECT counterpart --- 
D33J013343.4+304356.5. Two harmonics were fitted to 462 measurements  
(of a time span of 2615 days). \emph{Top panel} shows data of the DIRECT 1st 
campaign \citep{Macri01b}, in circles, of the DIRECT 2nd campaign \citep{Mochejska01a}, in crosses,
and data obtained here, in triangles. \emph{Middle panel} presents the periodogram 
where the 2nd strongest frequency is consistent with one half the strongest 
frequency value, respectively. \emph{Bottom panel} shows the 
combined phase-folded lightcurve, where the fitted zero point of each data set 
was subtracted. Fitted zero points 
are 19.43, 19.91 and 19.90 mag for the DIRECT 1st campaign 
(\citealt{Macri01b}, marked by circles), DIRECT 2nd campaign 
(\citealt{Mochejska01a}, marked by crosses) and the Wise data (marked by 
triangles), respectively. The black solid line is just an harmonic fit, derived 
by the lightcurves combining procedure.}
\label{fig:comblc} 
\end{figure}

Periodogram value for each frequency was taken to be the fitted 
amplitude divided by the ${\chi}^2$ goodness-of-fit parameter.
We then divided the whole periodogram by a fitted polynomial, using a cubic 
smoothing spline fit. 
Number of harmonics used for each frequency in the fit was determined by using the
Akaike Information Criterion \citep[AIC,][]{Akaike74}. 
Fig.~\ref{fig:comblc} shows an example of this analysis for W10821. 

Lightcurves of \combnumtotal periodic objects were combined with data 
obtained here. For \badcomb of those stars results
were unsatisfactory, mainly due to  
high noise level of faint objects.
Results of the combining analysis for \combnum objects are presented on 
Table~\ref{table:comb_lc}, together with the periods obtained previously
by the two DIRECT project programs. 

\begin{table*}
\begin{minipage}{13cm}
\caption{List of \combnum combined lightcurves.}
\label{table:comb_lc}
\begin{tabular}{rrr r@{.}l r@{.}l cc r@{.}l ccrr}
\hline
& & & \multicolumn{4}{c}{DIRECT} & & \multicolumn{6}{c}{Combined}\\
\hline
Desig. \ \ & & \ \ &\multicolumn{2}{c}{Period1}&\multicolumn{2}{c}{Period2}& & &\multicolumn{2}{c}{Period}& ${\Delta}$P& $\!\!$ Amp.& ZP1    & ZP2 \\
           & &     &\multicolumn{2}{c}{[days]} &\multicolumn{2}{c}{[days]} & & &\multicolumn{2}{c}{[days]}& [days]     & [mag]     & \ \ [mag]& \ \ [mag]\\
\hline

W21612&  & &       2&3372&  2&3362& & &  2&33727&  3.7e-04&    0.18&  0.06&  0.40 \\
W11486&  & &       2&7081&  2&7104& & &  2&70819&  6.9e-04&    0.11&  0.27& -0.64 \\
W10764&  & &       4&4330&  4&4363& & &  4&43312&  2.0e-04&    0.14&  0.07&  0.81 \\
W20942&  & &       5&0950&  5&0920& & &  5&09490&  1.6e-04&    0.27& -0.17&  0.71 \\
W21067&  & & \multicolumn{2}{c}{-}& 5&298&  & &  5&2618&   1.4e-03&    0.04& -0.03&  0.27 \\
W10304&  & &       8&14&    8&14&   & &  8&1367&   4.3e-03&    0.22&  0.19&  0.05 \\
W21568&  & & \multicolumn{2}{c}{-}& 4&869&  & &  9&78952&  9.7e-04&    0.04& -0.04&  0.18 \\
W10244&  & &      11&95&   11&97&   & & 11&9689&   8.1e-03&    0.34&  0.19& -0.29 \\
W11268&  & &      12&35&   12&36&   & & 12&353&    5.1e-02&    0.17&  0.11&  0.08 \\
W11572&  & &      12&92&   12&93&   & & 12&9199&   9.5e-03&    0.26& -0.06&  0.65 \\
W10128&  & &      13&04&   13&04&   & & 13&0634&   4.8e-03&    0.28&  0.03&  0.38 \\
W20935&  & &      13&31&   13&31&   & & 13&3156&   3.3e-03&    0.35& -0.11&  0.06 \\
W22343&  & &      13&56&   13&56&   & & 13&5685&   4.6e-03&    0.26&  0.18& -0.69 \\
W20873&  & &      14&35&   14&35&   & & 14&347&    1.6e-02&    0.33&  0.20& -0.51 \\
W11291&  & &      14&59&   14&61&   & & 14&5879&   7.2e-03&    0.21& -0.11& -0.09 \\
W10892&  & &      14&92&   14&93&   & & 14&925&    1.3e-02&    0.26&  0.00& -0.29 \\
W21428&  & &      15&08&   15&08&   & & 15&060&    3.0e-02&    0.28& -0.12& -0.47 \\
W20988&  & &      15&78&   15&78&   & & 15&785&    1.0e-02&    0.35& -0.07& -0.61 \\
W22150&  & &      16&28&   16&28&   & & 16&273&    6.1e-02&    0.34& -0.14& -0.21 \\
W20563&  & &      17&48&   17&48&   & & 17&50&     4.7e-01&    0.29&  0.19& -0.65 \\
W11733&  & &      17&83&   17&83&   & & 17&841&    3.3e-02&    0.39& -0.14&  0.27 \\
W10267&  & &      18&61&   18&60&   & & 18&6047&   7.9e-03&    0.55& -0.04&  0.06 \\
W11794&  & &      18&81&   18&81&   & & 18&797&    1.7e-02&    0.50& -0.04&  0.36 \\
W10821&  & &      20&19&   20&18&   & & 20&161&    3.2e-02&    0.26&  0.48&  0.47 \\
W21058&  & &  \multicolumn{2}{c}{-}&  \multicolumn{2}{c}{-}&    & & 20&161&    1.1e-02&    0.07& -0.03&  0.11 \\
W30811&  & &      20&18&   20&18&   & & 20&182&    1.8e-02&    0.24& -0.10&  0.02 \\
W10046&  & &      21&91&   21&93&   & & 21&954&    1.9e-02&    0.40&  0.02&  0.22 \\
W31015&  & &      22&67&   22&67&   & & 22&650&    2.1e-02&    0.36& -0.04&  0.01 \\
W20594&  & &      23&31&   23&31&   & & 23&310&    1.6e-02&    0.45&  0.03&  0.27 \\
W20138&  & &      24&88&   24&88&   & & 24&9066&   6.7e-03&    0.35& -0.10&  0.19 \\
W10805&  & &      26&39&   26&37&   & & 26&316&    2.7e-02&    0.25&  0.20& -0.23 \\
W10431&  & &      26&48&   26&48&   & & 26&490&    3.2e-02&    0.29&  0.09&  0.30 \\
W21275&  & &      26&50&   26&50&   & & 26&525&    4.4e-02&    0.30&  0.18&  0.13 \\
W10143&  & &      26&57&   26&58&   & & 26&596&    1.9e-02&    0.30&  0.06&  0.15 \\
W11552&  & &      27&28&   27&30&   & & 27&248&    1.8e-02&    0.43& -0.03&  0.06 \\
W10636&  & &      27&97&   28&00&   & & 28&011&    3.2e-02&    0.24&  0.13& -0.15 \\
W30926&  & &      30&29&   30&29&   & & 30&257&    3.4e-02&    0.35& -0.04& -0.28 \\
W10773&  & &      30&50&   30&52&   & & 30&488&    5.1e-02&    0.24&  0.16& -0.05 \\
W10367&  & &      31&25&   31&26&   & & 31&348&    2.0e-02&    0.46& -0.06& -0.03 \\
W10629&  & &      33&95&   33&98&   & & 34&014&    2.6e-02&    0.23&  0.13& -0.05 \\
W11386&  & &      35&86&   35&87&   & & 35&84&     1.5e-01&    0.40& -0.32&  0.17 \\
W21290&  & &      36&09&   36&09&   & & 36&036&    6.1e-02&    0.51& -0.28&  0.49 \\
W21743&  & &      36&35&   36&35&   & & 36&496&    4.8e-02&    0.19&  0.02& -0.12 \\
W20627&  & &      37&34&   37&34&   & & 37&453&    2.9e-02&    0.47&  0.06& -0.01 \\
W20702&  & &      55&91&   55&91&   & & 55&10&     1.5e-01&    0.33& -0.01& -0.17 \\
W10274&  & &      55&99&   55&79&   & & 55&25&     2.3e-01&    0.12&  0.00&  0.18 \\
W10466&  & &      57&45&   57&45&   & & 57&47&     1.2e-01&    0.16& -0.06& -0.18 \\
W20923&  & &      56&84&   56&84&   & & 69&44&     1.9e-01&    0.18& -0.06&  0.20 \\
W20280&  & &      73&94&   73&94&   & & 74&63&     7.5e-01&    0.29& -0.27&  0.46 \\
W10034&  & &      55&0&    54&930&  & &107&53&     6.3e-01&    0.10& -0.06&  0.31 \\

\hline

\end{tabular}

\medskip
Columns contain (1) Wise Designation, (2) period of \cite{Macri01b}, 
(3) period of \cite{Mochejska01a,Mochejska01b}, (4) period of the combined 
lightcurves, (5) Period uncertainty, (6) amplitude of the fit, 
(7) zero point shift of \cite{Mochejska01a,Mochejska01b} relative to \cite{Macri01b},
and (8) zero point shift of the Wise data relative to \cite{Macri01b}.

\end{minipage}
\end{table*}

Except for two objects, W10034 \& W20923, period derived 
by the combining procedure is consistent
with those derived by analyzing each data set separately. 
For W10034, analysis of the combined lightcurve was able to 
reveal the true period (107.53 days), which is twice the period derived 
previously.
For W20923, we have detected a period of 69.44 days, while the DIRECT
period is 56.84. It is also interesting to note that a period of 69.50 days 
was already derived by \citet[][Table~II, variable No. 10]{Hubble26} for 
this object.

\begin{figure}
\centering
\includegraphics[height=7.5cm,width=10cm]{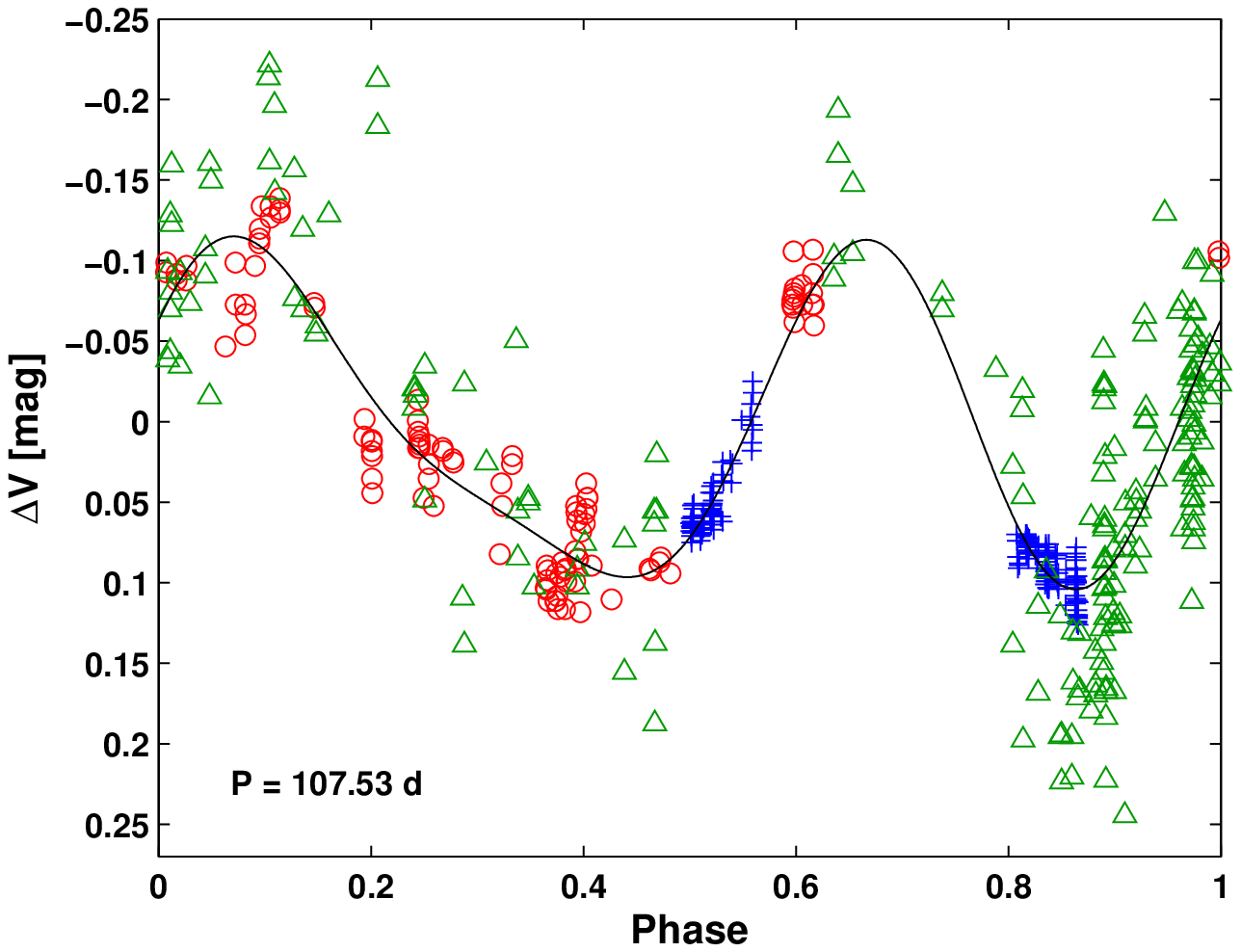}
\caption{Combined lightcurve of W10034 with its DIRECT counterpart --- D33J013420.8+303943.0, 
revealing the true period of 107.53 days, twice then period derived previously.
Only by analyzing the combined lightcurve, consisting of 490 measurements taken in a time span 
of 2608 days, the true period was derived. 
The different markings for each data set are the same as in 
Fig.~\ref{fig:comblc}, bottom panel.}
\label{fig:10034_comb}
\end{figure}

Table~\ref{table:complist} lists all variables detected here along with 
designation of known variables. 

\begin{table*}
\begin{minipage}{150mm}
\caption{List of all \allvar variables detected in this survey. Within each 
variability type objects are sorted by increasing period and the non-periodic 
variables are sorted by increasing R.A. For previously known variables their known 
designation is given. }
\label{table:complist}

\begin{tabular}{cccclclclc}

\hline
\multicolumn{5}{c}{Wise}&& \multicolumn{4}{c}{Known} \\
Desig. & R.A.& Dec.& Type& Period& &Designation& Type& Period& Comments\\ 
   & hh mm ss.ss & dd mm ss.s& &[days] & &&  & [days]& \\
\hline

W21612& 01 33 56.26& +30 33 58.8& E&  2.337324&& D33J013356.2+303358.6&	E&   2.34   &\\
W11486& 01 33 54.32& +30 40 28.9& E&   2.70811&& D33J013354.3+304029.6&	E&   2.71   &\\
\\
W11572& 01 33 45.93& +30 42 30.3& C&    12.9199&& D33J013346.0+304231.9&	C&  12.93& 10\\
W10128& 01 34 02.76& +30 41 44.4& C&    13.0634&& D33J013402.8+304145.7&	C&  13.04& 10\\
\\
W21162& 01 33 60.96& +30 33 54.8& P&    2.1142&& D33J013359.8+303354.9&	P&   2.12   &\\
W21067& 01 33 41.63& +30 32 20.5& P&    5.2618&& D33J013341.6+303220.3&	P&   5.30   &10\\
\\
W31624& 01 32 55.69& +30 35 34.8& N& -	     &&	-	     &	-&   -	    &1,4\\
W31074& 01 32 56.73& +30 39 04.0& N& -	     &&	-            &	-&   -	    &1,4\\

\vdots&   \vdots&       \vdots& \vdots& \vdots&&  \vdots&      \vdots& \vdots& \vdots\\
\hline

\end{tabular}

\medskip
Key to variability types: E --- Eclipsing binary, C --- Cepheid, P --- 
Unclassified Periodic, N --- Non-periodic variable (DIRECT miscellaneous 
type). \\
Comments: 1: New variable; 2: New classification of a known variable; 
3: Position in DIRECT M33C, reported in Table~9 of Macri et al. (2001b); 
4: Position in DIRECT M33C (new variable); 5: Position outside all DIRECT 
fields (new variable); 6: Object detected as periodic variable by 
Shemmer et al.\ (2000); 7: Astrometric matching between Wise and DIRECT 
objects can not be determined accurately due to the crowded region; 
8: classified as a Long Period Variable 
(LPV) by Kinman, Mould \& Wood (1987); 9: Object detected as variable by 
Hubble \& Sandage (1953); 10: Period derived from the combined 
lightcurve.\\
Only a small sample of Table~\ref{table:complist} is presented here, 
including two objects from each variability type. The entire table is 
available in the MNRAS electronic issue.  
\end{minipage}
\end{table*}

\subsection{Comparison with the DIRECT results}
\label{subsec:comp_dir}

\begin{table*}
\begin{minipage}{140mm}
\caption{Summary of the detections by the Wise Observatory}
\label{table:summary}
\begin{tabular}{lccccccccc}
\hline
Var. Type     & All Fields \ \ & \multicolumn{2}{c}{M33A \& M33B\ \ }  & \multicolumn{2}{c}{M33C\ \ } & 
Out & Previous& New Class.& New Var.\\
&             &Wise   &  DIR   & Wise   &  DIR &      &    &    &         \\  
\hline    
\hline    
Cepheid       &$\ \ $45  &$\ \ $34      &$\ \ $34     &11       &$\ \,$6  &-   &-   &$\ \ \ $3$^{a}$    &$\ \ \ \,$5\\
EB            &$\ \ $12  &$\ \ \ \, $8  &$\ \ \ \, $7 &$\ \, $4 &$\ \,$1  &-   &-   &$\ \ \ $1$^{b}$    &$\ \ \ \,$4\\
Unclass. Per. &$\ \ $56  &$\ \ $40      &$\ \ $34     &16       &$\ \,$1  &-   &2   &$\ \ \ $7$^{a}$    &$\ \ $19\\
Non Per.      &177       &136           &$\ \ $73     &38       &$\ \,$2  &3   &3   &-          &$\ \ $99\\
\hline    
Total         &290       &218           &148          &69       &10       &3   &5   &11         &127\\
\hline
\end{tabular}

\medskip
Columns contain: (1) variability type and (2) number of detected variables. 
(3) Variables detected here inside the DIRECT M33A and M33B fields and 
(4) of those, number of variables detected by the 
DIRECT, including objects for which a new variability class was obtained here.
(5) Variables detected here inside the DIRECT M33C field and (6) of those, 
number of variables detected by the DIRECT (\citealt{Macri01b}, Table~9). 
Only variables in M33C and \emph{not} in overlapping regions are listed on 
columns 5 and 6. (7) Variables detected outside the DIRECT FOVs. (8) Variables 
detected here, detected previously by studies other than the DIRECT. 
(\citealt{Hubble53}, \citealt{Kinman87} and \citealt{Shemmer00}). (9) Newly 
classified variables and (10) new variables.\\
Notes: (a) classified by DIRECT as non-periodic. (b) classified 
by DIRECT as an unclassified periodic.
\end{minipage}
\end{table*}

Many objects classified here as variables were not classified as such 
by the DIRECT project. Out of the total 290 variables detected here 
only 158 variable objects were astrometrically 
matched to DIRECT variables. The other 132 variables consist of
102 non-periodic and 30 periodic variables. Of those 132, 5 were already 
identified as variables by previous studies (\citealt{Hubble53}, 
\citealt{Kinman87} and \citealt{Shemmer00}), giving a total of 
127 new variables, consisting of 99 non-periodic and 28 periodic.
In addition, we obtained an improved variability type 
for 11 of the 158 astrometrically matched variables. 
 
The M33 area monitored here includes three fields, while the DIRECT 
catalogs of \citet{Macri01b} and \citet{Mochejska01a,Mochejska01b} list 
variables from only two fields, M33A and M33B. Those fields are also slightly 
smaller than the corresponding \emph{direct1} and \emph{direct2} fields 
observed here. 
Considering only the 221 variables we find here 
inside the FOV of DIRECT M33A and M33B fields, 
148 of them are present in the variables catalogs of \citet{Macri01b} or 
\citet{Mochejska01a,Mochejska01b}. The other 73 variable objects 
include 63 non-periodic and 10 periodic variables. 

Table~\ref{table:summary} lists 
the number of variables detected here, of each variability type, 
and of those, number of variables detected by the DIRECT project. 
Table~\ref{table:summary} also refers to variables positioned in the 
DIRECT M33C field since although this field's variables were not reported by 
the DIRECT, 15 of them are included in Table~9 of \citet{Macri01b}.
Also given in Table~\ref{table:summary} are number of variables detected outside
all three DIRECT fields, variables detected by other, previous studies, number
of newly classified variables and of new variables. 

\section{Interesting Variables}
\label{sec:int_var} 

\subsection{W21067 --- An Optical Periodic X-Ray Source}
\label{subsec:21067} 

\begin{figure}
\centering
\includegraphics[height=8cm,width=10cm]{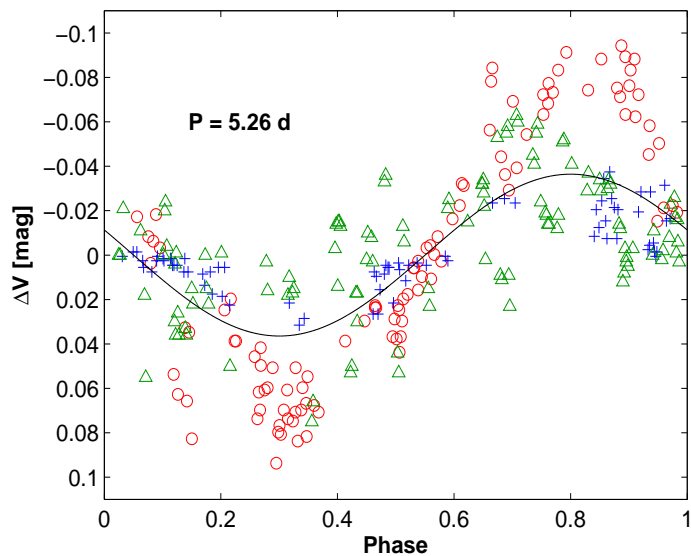}
\caption{Combined lightcurve of W21067 folded on the period detected by the 
combining procedure, of 5.2618 days. Data of DIRECT 1st campaign 
\citep{Macri01b} is marked in circles, of DIRECT 2nd campaign 
\citep{Mochejska01b} in crosses and data obtained here in triangles. 
The solid line is a single harmonic fit calculated by the combining 
procedure (See section~\ref{subsec:comb_lc}).}
\label{fig:21067_lc}
\end{figure}

The combined data of W21067 (1:33:41.63, +30:32:20.5), one of the
brightest objects in the sample, yielded a period of 5.26 days with an
amplitude of 0.04 (see Fig.~\ref{fig:21067_lc}). 
An X-ray source at a distance of 1.4'' from our position of W21067 was
detected by \citet[Table~3, source 194]{Pietsch04} using the
XMM-Newton observatory. They have measured an 0.2--4.5 KeV flux of
$(1.3900 \pm 0.0097)\times10^{-14}$ erg/sec/cm$^{2}$ and also
suggested that the positional correlation between an optically
periodic variable and an X-ray source makes this object an XRB
candidate.

\begin{figure}
\centering
\includegraphics[height=9cm,width=17cm]{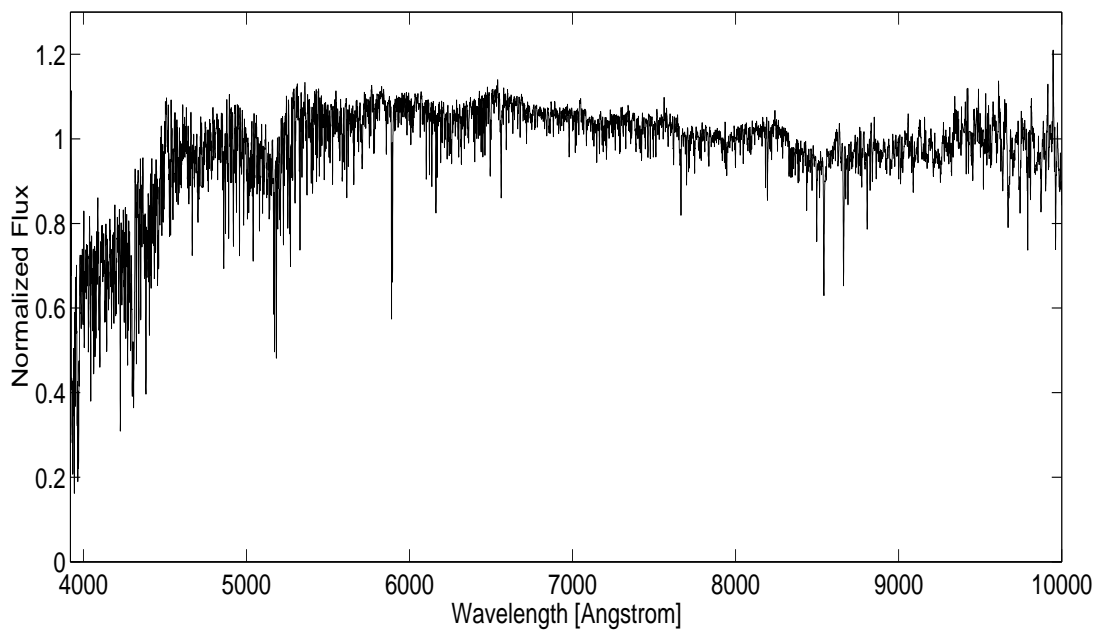}
\caption{Joined spectrum from all 10 orders of W21067, taken by the 
Keck II ESI. Normalized flux is plotted against wavelength. 
Spectrum was smoothed using a running mean of 5 adjacent pixels.}
\label{fig:21067_spec_joined}
\end{figure}

To follow this suggestion Alceste Bonanos of the CfA
obtained for us a multi-order spectrum of W21067 with the Echellette
Spectrograph and Imager (ESI) at the Keck II 10-m telescope in the
Echelle mode. Fig.~\ref{fig:21067_spec_joined} presents the joined
spectrum of all 10 orders (orders 6 to 15), with a complete spectral
coverage from 3900--10000~\ang$\!\!$.  Velocity dispersion is about
11.4 km/sec/pixel in all orders.

By correlating W21067's spectrum with the ELODIE library of template
spectra we were able to determine it is a K star, with a
heliocentric radial velocity of $+16.4\pm0.6$ km/s. The derived velocity
is significantly different from the systemic radial velocity of M33, 
of $-179\pm3$ km/s \citep{DeVaucouleurs91}.  
The most plausible conclusion therefore is that W21067 is \emph{not} an 
M33 member but a \emph{Galactic} foreground object.

The many absorption lines and absence of significant emission features in 
W21067 spectrum rules out the possibility of a compact binary companion  
as the X-rays source \citep{Bradt83}. Therefore we propose 
chromospheric activity as the source of the optical periodicity 
and X-ray radiation. 

The periodic modulations of chromospherically active stars are known
to show long-term amplitude and shape variations, resulting from their
dark regions evolution during solar-like activity cycles
(\citealt{Guinan92}, \citealt{Olah00}, \citealt{Rodono92}).  This
might be the cause for the somewhat large scatter in W21067 combined
folded lightcurve, consisting of data from a period of $\sim$7 years.

\begin{figure}
\centering
\includegraphics[height=10cm,width=10cm]{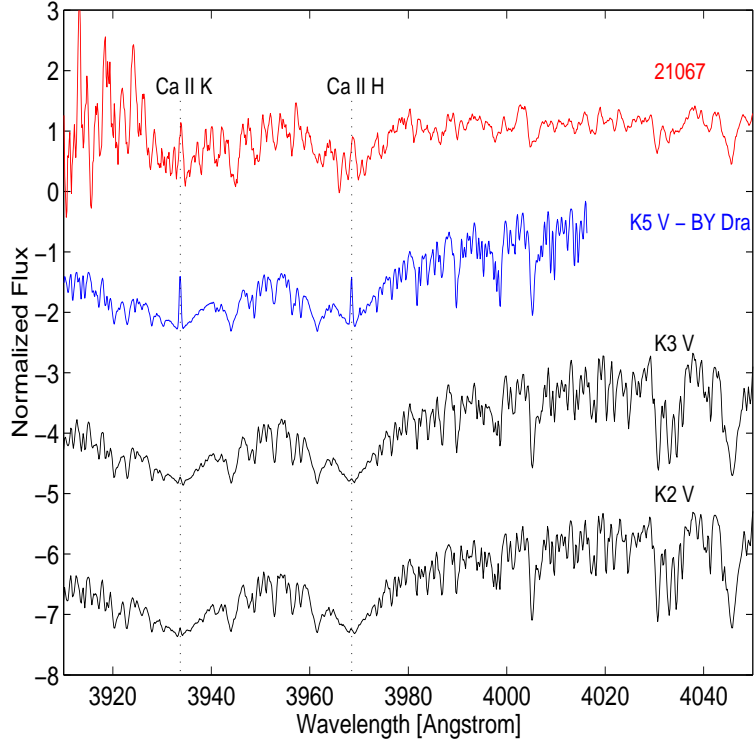}
\caption{From top to bottom: Spectrum of W21067 at the Ca II H \& K spectral region. 
Spectrum of a chromospherically active star, HD 201091, a BY Dra K5 V star. 
Spectra of two inactive stars for comparison
(HD 4628 K2 V, HD 16160 K3 V). Spectra of comparison stars were taken from 
\citet{Montes97}.}
\label{fig:21067_spec_comp2}
\end{figure}

Fill-in cores of the Ca II H \& K absorption lines, at 3968.5 \ang and 3933.7
\ang$\!\!$, respectively, are a classical spectral feature used to
identify chromospheric activity (\citealt{Fekel93}, \citealt{Strassmeier93}).
Fig.~\ref{fig:21067_spec_comp2} shows W21067 Ca II H \& K lines
together with an active star and two inactive stars of a similar
spectral type. Despite the decreased S/N close to the spectrum's blue
end caused by the lower CCD QE and increased atmospheric extinction
at these wavelengths, fill-in cores of W21067 Ca II H \& K lines can be
noticed.

The color-temperature color-index relation \citep{Allen73} for the
DIRECT $B - V= 0.79~\pm~0.04$ color index yields a color temperature
 of \ $T_{c}~=~5250~\pm~150^{\degt}$K. For Galactic
object in the direction of M33 extinction and reddening can be
neglected, and therefore we can assume the temperature estimate is
valid.  The effective temperature bolometric correction (BC) relation
\citep{Allen73} yields $BC~=~0.18~\pm~0.03$ mag, which results in an
X-ray to $V$ flux ratio of $log(f_X/f_{V})~=~-2.72\pm0.04$, consistent
with the mean value of $-2.9~\pm~0.8$ (median $-2.8$), given by
\cite{Padmakar00} for 202 active binaries.

\subsection{W31230 --- An Optical Periodic at SNR Position}
\label{subsec:SNR} 

\begin{figure}
\centering
\includegraphics[height=8cm,width=10cm]{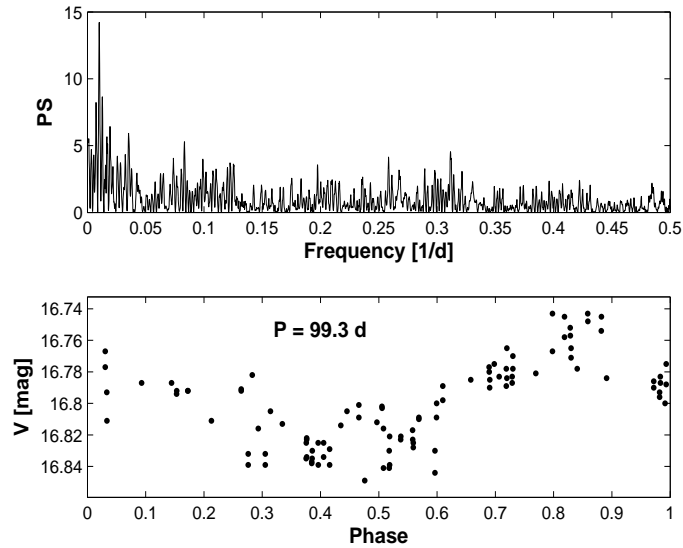}
\caption{\emph{Top panel} presents a Lomb-Scargle power spectrum of W31230
(normalized to unit mean) and the \emph{bottom panel} shows the phased 
lightcurve, with an amplitude of 0.04 mag.}
\label{fig:31230}
\end{figure}

W31230 (1:33:11.17, 30:34:21.9) is located 0.41'' from SNR 19 
of \cite{Gordon98} and was detected here as an unclassified periodic 
variable. Considering the number of SNRs and \emph{periodic} variables 
detected here in the direct3 field, the probability of a random positional 
correlation is 0.0009. 
\cite{Calzetti95} list this object as an $H_{\alpha}$ emitter 
and \cite{Sholukhova99} detected an $H_{\alpha}$ emitter at a distance of 2.2'' from 
W31230, for which they measured an $H_{\alpha}$ equivalent width of 19 \AA. W31230 
is located at the DIRECT M33C field and was not reported as variable by them. 
Our analysis detected a period of $99.3\pm4.4$ days and an amplitude
of 0.04 mag. Fig.~\ref{fig:31230} presents a Lomb-Scargle \citep{Scargle82} 
power spectrum and a phased lightcurve of our data for this star. 

\subsection{Optical Variables at Wolf-Rayet Positions}
\label{subsec:WR} 

WR Variability can originate from random winds variations, rotation,
spots evolution and pulsation, radial and non-radial
\citep{Moffat86}. A persistent periodic variability might indicate a
WR in a binary system where variability is induced by either geometric
eclipses, wind eclipses or proximity effects \citep{Marchenko98}.

By correlating our variables against the SIMBAD database we found two
objects which are located at positions of WR stars and a third object
located at the position of a WR candidate. All three astrometric
matches are within 1.1''.

\textbf{W20549} (1:33:58.67, 30:35:26.6), a spectroscopically
confirmed M33 WR \citep[WR115]{Massey98} of type WNE \citep{Massey95},
was identified here as a new EB, with a period of 24.4920 days and orbital 
eccentricity of 0.25 
(see Fig.~\ref{fig:eblc} and Tables~\ref{table:neweb} \& \ref{table:ebas}).
\textbf{W31347} (1:33:27.25, 30:39:09.7), a spectroscopically
confirmed M33 WR of type WN9 \citep[WR39]{Massey98}, was detected as a
new non-periodic variable (see Fig.~\ref{fig:varlc} and
Table~\ref{table:newvar}).  \textbf{W10975} (1:34:20.30, 30:45:45.4)
was identified as a WR candidate by \citet[Table~2, object
29]{Massey87} and was detected here as a new non-periodic variable
(see Fig.~\ref{fig:varlc} and Table~\ref{table:newvar}).

\subsection{W31284 - M33 Variable C}
\label{subsec:31284}

W31284 (1:33:35.13, 30:36:00.8) is a known LBV, named M33 Variable C, 
discovered by \citet[Fig.~6]{Hubble53} by using data from 1921 to 1953. 
Since then, this star was included in a few other long-term surveys:
\citet[Fig.~8]{Rosino73}, from 1960--1972, \citet[Fig.~5]{Kinman87} 
from 1982--1985 and \citet[Fig.~2]{Kurtev99} from 1982--1990.
   
\begin{figure}
\centering
\includegraphics[height=8cm,width=10cm]{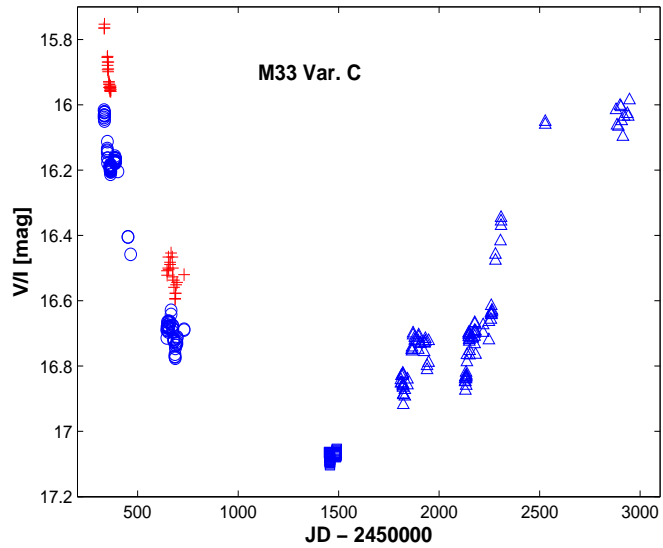}
\caption{M33 Var. C lightcurve from DIRECT and this work. $V$-band data of 
the DIRECT first campaign is presented in circles, of the DIRECT second 
campaign in squares and from this work in triangles. DIRECT first campaign 
$I$-band data is marked by crosses. $V$-band lightcurve shows a decrease of 
about 1 magnitude in the star's luminosity followed immediately by a similar 
increase. 
Variability time-scale is shorter than observed by Hubble \& Sandage (1953) 
and Rosino and Bianchini (1973). }
\label{fig:31284}
\end{figure}

In both DIRECT campaigns and in this work M33 Var. C was identified as 
a non-periodic variable. Lightcurve containing data of the three telescopes, 
with a time span of 7.16 years, is presented on Fig.~\ref{fig:31284}. 
Although there might be small zero-point shifts between telescopes, 
a decrease of 
more than 1 magnitude, followed by a similar increase is evident. 
This lightcurve and the one presented by \cite{Kinman87}, show 
brightness variations with a much shorter time-scale than observed by 
\cite{Hubble53} and \cite{Rosino73}. Furthermore,  
Fig.~\ref{fig:31284} shows that the star stayed at minimum brightness for a 
relatively short time before brightening again, a behavior which was not 
observed previously.
This suggests that M33 Var. C brightness variations became more rapid since
the 1980s. 

\subsection{W11324 - LBV Periodic Microvariability Confirmation}
\label{subsec:11324} 

W11324 (1:34:06.59, 30:41:46.8) is known as B416 \citep{Humphreys80}. 
It was identified by \citet*[hereafter SLS]{Shemmer00}
as an LBV, confirming \cite{Massey96} LBV candidate classification.
SLS also report a periodic
microvariability of 8.26 days with a $V$ amplitude of 27 mmag, 
a $V$ magnitude of 16.7$\pm$0.1 and
an $H_{\alpha}$ EW of 106$\pm$2 \ang$\!\!$, consistent with a $V$ magnitude of
16.76 and $H_{\alpha}$ EW of 109.1 \ang$\!\!$, measured by \cite{Calzetti95}.
W11324 was not classified by DIRECT as a variable.
Here, W11324 was detected as an unclassified periodic, with a period of 
8.251$\pm$0.017 days. The consistency of our period and SLS period, derived 
from observations taken more than 10 years apart, confirms the periodic nature 
of this object. 

\subsection{W10913 - Hubble's Variable 19 in M33}
\label{subsec:10913}

W10913 (1:33:57.02, 30:45:11.5) was identified as variable by 
\citet[V19]{Hubble26} who classified it as a 54.7 days Cepheid. 
\cite{Macri01c} have shown that this object has undergone a dramatic 
decrease in its brightness variation, from a $B$ amplitude of 0.55 mag 
in Hubble's lightcurve to no detectable variation in the DIRECT data. 
In our data, this object is also 
non-variable and has a $V$ RMS of 0.03 mag, similar to the 
scatter of other non-variable stars of the same brightness.

\section{Conclusions}
\label{sec:conclusions}

We presented here the results of a long-term monitoring of the bright stars in M33, complete to  $V \sim 19.5$~mag. We discovered 8 new Cepheids, 5 binaries, 26 unclassified periodic variables with period range up to almost 300 days, and 99 non-periodic variables. Combined with the publicly available data of the DIRECT project, our data covers more than 5 years of stellar variability. One of these variables is the famous M33 Var. C, which displays more than 1 mag modulation with a timescale of 2500 days. The variability found here is different from that previously reported and therefore further observations of this star will be most interesting.

One intriguing result of this study is the findings about W31230, one
of the brightest stars in M33, which is found at the location of SNR 19 
of \cite{Gordon98} and shows $H_{\alpha}$ emission. It displays
a small periodic modulation of 0.04 mag at a period of about 100
days. This is similar to one of the two periodic modulations found for
the famous Galactic source SS 433 \citep{Margon84}, which is also
located at the center of Galactic SNR (e.g., \citealt{Kirshner80}) and
has a prominent $H_{\alpha}$ emission. The 165 d periodicity of SS 433
is associated with the precessing disc around the compact object,
which is probably the remnant of the SN explosion. SS 433 displays
another periodicity of 13 days (e.g., \citealt{Mazeh87}), associated
with the binary period (e.g., \citealt{Cherepashchuk81}). We could not
find a shorter periodicity at the W31230 data. If W31230 is indeed
similar to SS 433, the lack of the short periodicity could be caused
by a smaller inclination angle in our case. It would be therefore of
special interest to obtain a spectrum of this interesting star, to see
if there is any resemblance to the special spectral features of SS 433
\citep{Margon84}. 

Followup spectroscopy turned out another
interesting candidate, W21067, to be a foreground Galactic
chromospheric star. We therefore suggest that the interesting cases
found in M33 by photometry should all be followed by
spectroscopy. With multi-object spectrographs available today on a few
telescopes, such followup observations would require a relatively small 
amount of telescope time.

\section*{acknowledgments} We are indebted to the many observers of the Wise 
Observatory for gathering the data. We also wish to thank the 
Wise Observatory staff: E. Mashal, F. Loinger, S. Ben-Guigui and J. Dan. 
We have benefited from the assistance and fruitful conversations with 
K. Stanek, S. Kaspi, S. Zucker and E. O. Ofek. 
We would like to express our deep gratitude to A. Bonanos for obtaining for us W21067 spectra.
We thank M. Mayor and S. Udry for using their library of spectra to derive 
W21067 radial velocity. We wish to thank the referee for his thorough 
reading of the paper and for his very useful suggestions.
This work was partially funded by the German-Israeli Foundation for Scientific
Research and Development and by the Israeli Science Foundation. 
This research has made use of NASA's Astrophysics Data System Abstract
Service and of the SIMBAD database, operated at CDS, Strasbourg, France.  

\bibliography{mybib}

\label{lastpage}

\end{document}